\documentclass{aa}
\usepackage{amsmath}

\usepackage{graphicx,txfonts,natbib}
\usepackage{longtable}

\begin{document}
\title{Evidence for nonlinear resonant mode coupling in the $\beta$ Cep star
HD\,180642 (V1449 Aql) from CoRoT space-based photometry\thanks{The CoRoT space
mission was developed and is operated by the French space agency CNES, with
participation of ESA's RSSD and Science Programmes, Austria, Belgium, Brazil,
Germany, and Spain.}}

\author{P.~Degroote\inst{1} 
\and M.~Briquet\inst{1}\thanks{Postdoctoral Fellow of the Fund for Scientific
Research, Flanders} 
\and C.~Catala\inst{2}
\and K.~Uytterhoeven\inst{1,3,4}
\and K.~Lefever\inst{1,5}
\and T.~Morel\inst{1,6}
\and C.~Aerts\inst{1,7}
\and F.~Carrier\inst{1}
\and M.~Auvergne\inst{2}
\and A.~Baglin\inst{2}
\and E.~Michel\inst{2}
\offprints{P. Degroote}}

\institute{Instituut voor Sterrenkunde, K.U.Leuven, Celestijnenlaan 200D, B-3001
Leuven, Belgium 
\and LESIA, CNRS, Universit\'e Pierre et Marie Curie, Universit\'e Denis
Diderot, Observatoire de Paris, 92195 Meudon cedex, France
\and INAF-Osservatorio Astronomico di Brera, Via E. Bianchi 46, 23807 Merate,
Italy 
\and
Laboratoire AIM, CEA/DSM-CNRS-Universit\'e Paris Diderot; CEA, IRFU, SAp, centre
de Saclay, F-91191, Gif-sur-Yvette, France
\and
Belgisch Instituut voor Ruimte Aeronomie (BIRA), Ringlaan 3, B-1180 Brussels,
Belgium 
\and
Institut d'Astrophysique et de G\'eophysique, Universit\'e de Li\`ege, All\'ee
du 6 Ao\^ut 17, B-4000 Li\`ege, Belgium
\and Department of Astrophysics, IMAPP, University of Nijmegen, PO Box 9010,
6500 GL Nijmegen, The Netherlands}

\date{Received 2 March 2009; accepted 27 May 2009}
\authorrunning{Degroote et al.}
\titlerunning{The CoRoT light curve of the $\beta\,$Cep star HD180642}

\abstract{We present the CoRoT light curve of the $\beta\,$Cep star HD\,180642,
assembled during the first long run of the space mission, as well as archival
single-band photometry.}  {Our goal is to analyse the detailed behaviour present
  in the
light curve and interpret it in terms of excited mode frequencies.}  {After describing
the noise properties in detail, we use
various time series analysis and fitting
techniques to model the CoRoT light curve, for various 
physical assumptions. We apply 
statistical goodness-of-fit criteria that allow us to select the
most appropriate physical model fit to the data.}  
{We conclude that
the light curve model based on nonlinear resonant frequency and phase locking provides the
best representation of the data. The interpretation of the residuals is dependent on
the chosen physical model used to prewhiten the data.}  {Our observational results constitute a
fruitful starting point for detailed seismic stellar modelling of this
large-amplitude and evolved $\beta\,$Cep star.}

\maketitle

%

\section{Introduction}
The B1.5II-III star HD\,180642 (variable star name V1449 Aql, $V_{\rm
mag}=8.29$) was identified as a candidate new $\beta$ Cep star by
\citet{waelkens1998} from Hipparcos data. This classification was confirmed by
\citet{aerts2000}, who identified the detected frequency of 5.4871\,d$^{-1}$
(63.508$\mu$Hz) as a radial mode with a large amplitude of 39\,mmag in the V
band, by interpreting amplitude ratios derived from multicolour Geneva
photometry obtained with the P7 photomultiplier instrument attached to the 0.70m
Swiss telescope at La Silla, Chile.

Given that HD\,180642 is the only known $\beta$ Cep star with appropriate
magnitude in the field-of-view of the CoRoT space mission \citep[see ``The CoRoT
Book'';][]{fridlund2006}, we undertook a preparatory observing effort to
assemble data to be added to the CoRoT light curve. Several
high-resolution spectra were taken with the FEROS@2.2-m ESO/MPI telescope in La
Silla, Chile, in 2005.  The latter led to an estimate of the fundamental
parameters of the star: $T_{\rm eff}=24\,500\pm 1\,000$K, $\log g=3.45\pm 0.15$,
as well as an overall line broadening of 44\,km\,s$^{-1}$ and some evidence for
a mild nitrogen excess \citep{morelaerts}, as discovered in other $\beta\,$Cep
stars from high-precision spectroscopy \citep{morel2006,morel2008}, which could
betray the existence of deep mixing.  The combination of the low gravity and
high pulsational amplitude of this class member is rather exceptional (see
\citet{stankov2005}, their Fig.\,8) and seems to suggest an
object near the end of the core-hydrogen burning phase, almost ready to cross the
Hertzsprung gap in the Hertzsprung-Russell diagram.

Adding the space photometry from Hipparcos to CoRoT and ground-based data, brings the total timespan
of observations to 18 years. The dominant mode of the star is present in all of these datasets and we thus have the means to determine its frequency stability over time.  On the other hand,
the high time-sampling of the CoRoT light curve combined with its low noise
level, give us the possibility to look for variability far beyond this dominant
mode. The richness of the CoRoT frequency spectrum led at once to the conclusion
that the monoperiodicity of the star must be refuted, as was already suggested
by \citet{uytterhoeven2008} from the ground-based data.

We have entered a new stage of precision in observational astronomy with the
CoRoT space mission.  In this paper we thoroughly investigate the variability of
HD\,180642 from single-channel photometry. Additional time series of multicolour photometry and
high-resolution spectroscopy of the star are the subject of a twin
paper \citep{briquet2009}.

\section{Observations}

\subsection{The CoRoT data }

\begin{figure*}
\includegraphics[width=2\columnwidth]{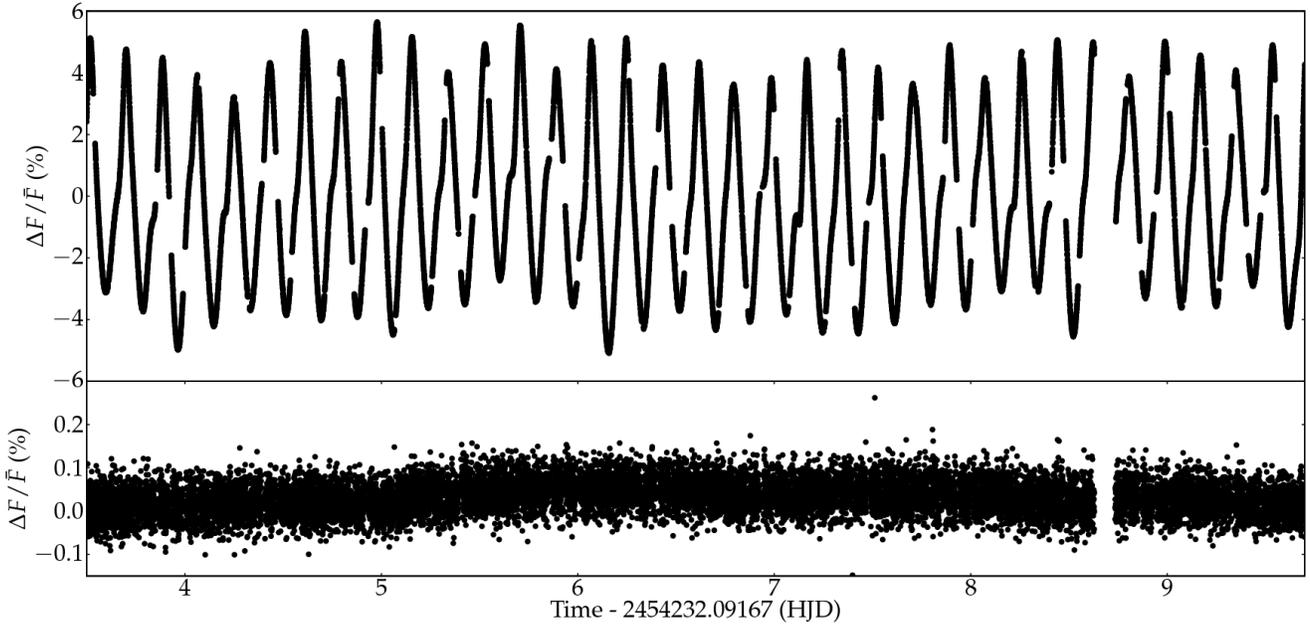}
\caption{Part of the reduced CoRoT light curve (upper panel), containing 379\,785 datapoints
in total. Despite the intrinsic equidistant nature of space based observations,
gaps are unavoidably present, mainly due to the regular passage of the
satellite through the South Atlantic Anomaly \citep{auvergne2009}. The lower panel shows the CoRoT
light curve of the constant star HD\,181072.}\label{fig:reduced_data}
\end{figure*}

The raw light curve from the CoRoT database contains 422\,949 datapoints, with
an average time sampling of 32\,s during 156.6~days and starting on $t_0={\rm
HJD}\ 2454232.091674$. This brings the Nyquist frequency up to
1350\,d$^{-1}$. To obtain the highest possible precision, roughly $10\%$ of the
datapoints were deleted because of flagged datapoints ($9.8\%$) and extreme
outliers ($0.5\%$ have an estimated error value above the $6$\,$\sigma$ level),
keeping 379\,785 datapoints (Fig.~\ref{fig:reduced_data}).

\begin{figure}
\includegraphics[width=\columnwidth]{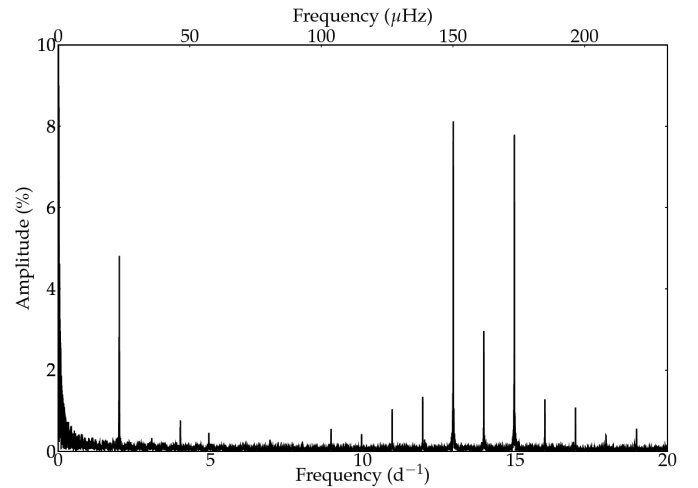}
\caption{Spectral window of the CoRoT measurements of HD180642 after removing
flagged datapoints and outliers. The highest peak is 8\% of the main peak,
and is well separated from it.}\label{fig:window_function}
\end{figure}

After rigorous tests, we decided not to interpolate the remaining points,
because the improvements of the spectral window do not weigh up against the
introduced uncertainties connected with the gap filling model. The highest
amplitude in the window function is only $\sim$\,$8\%$ of the main amplitude
(Fig.\,\ref{fig:window_function}). This means that we effectively spread out the
power of each peak over several peaks, mainly well separated by
$\sim$\,$13$\,d$^{-1}$.

A raw estimate of the noise level of the light curve, computed
as the average of the periodogram between 30\,d$^{-1}$ and 40\,d$^{-1}$ is at
57\,$\mu$mag or 0.00536\,\% in relative flux units, and slowly decays at higher
frequencies to 24\,$\mu$mag or 0.0026\,\% between 100\,d$^{-1}$ and
110\,d$^{-1}$. We do not convert the light curve to magnitudes because the
transformation from flux is not uniform and the theoretically predicted
variations in first order are only linear in flux units. An exception is made in
the case where the CoRoT light curve is used in combination with ground based
observations. This does not pose a problem because of the dominant mode's large
amplitude.

The final reduced version of the light curve has also been corrected for long
term trends: among an
exponential, parabolic and linear trend, the linear trend resulted in the best
fit, reducing significantly the power in the periodogram at low frequencies. An instrumental cause of the trend seems most probable,
although long term (periodical) variations in the brightness of the star cannot
be excluded from this time series alone. Previous ground based observations
disfavour the second possibility, but do not exclude it, however.

\subsection{The noise properties of the CoRoT data\label{sect:noise}}

The CoRoT data products contain information on the standard deviation of the
star's intensity per second, which is interpreted as the noise on the data. For the use and
interpretation of the applied data analysis tools, it is vital to have a good understanding of the
noise properties. We divided the error by the local average flux value
of the light curve represented by a polynomial fit. A comparison between the Fourier periodogram of the intensity measurements
and the derived standard deviations, shows that in the case
of HD\,180642, the latter are contaminated by the former, and are thus not a reliable estimate of the instrumental noise. Instead, we use the
measurements of the constant star HD\,181072 of spectral type A2 and visual magnitude of 9.14, which has been observed simultaneously on the same
 CCD, as an appropriate representation of the noise properties.

Traditionally, uncorrelated homoscedastic white Gaussian noise is assumed in frequency analyses of $\beta$\,Cep stars. If the number
of data points is large enough, this implies that the distribution
of the normalised Fourier periodogram can be reasonably well approximated by an exponential distribution \citep{schwarzenberg1998}.
In the following paragraphs, we show that none
of the assumptions are strictly true, but the deviations are so small that the classical methods can still be applied provided that
a correction for correlated data is used.

\begin{figure}
\includegraphics[width=\columnwidth]{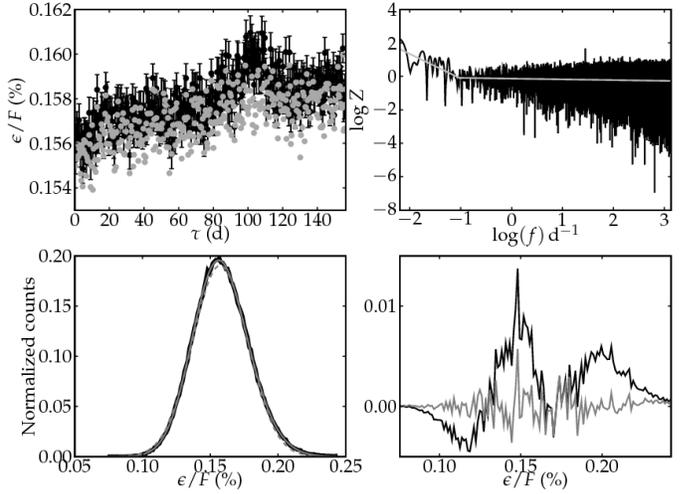}
\caption{Basic properties of the noise on the data: (\emph{upper left}) average (black) and median (grey) per sample of 1000 points. A small temporary bump and a continuously rising trend are visible. The fact that the median is consistently lower than the average, suggest a nonzero skewness. (\emph{upper right}) $\log-\log$ plot of the Scargle periodogram, gray lines are straight line fits. The noise is mostly white, except for a low frequency steep decay due to correlation effects. (\emph{lower left}) Histogram of the data (black) with a normal fit (dashed grey line) and a skewnormal fit (solid grey line). The skewnormal distribution fits the wings better than the normal distribution. (\emph{lower right}) The residuals of the histogram fits, after subtracting a normal fit (black) and after subtracting a skew normal fit (grey), show that a skew normal distribution is a better estimation of the overall noise distribution }\label{fig:noise_properties}
\end{figure}

By binning the noise
measurements in samples of 1000 points, we can see that the noise is not uncorrelated nor homoscedastic:
we identify a continuously rising trend of $(1.71\pm0.3)\times 10^{-5}$\,percent\,d$^{-1}$
and a small temporary bump around day 130 (Fig.\,\ref{fig:noise_properties}). From a $\log$-$\log$ plot of the Scargle periodogram, it is apparent
that the noise is not white: at very low frequencies ($<0.1$\,d$^{-1}$) there is some
power excess due to the correlation effect. However, white noise is a good enough
approximation for $f>0.1$\,d$^{-1}$.

Drawing random samples of 1000 points, reveals that the noise is also not Gaussian: the sample mean is consistently
higher than the sample median, suggesting that the noise has nonzero skewness. When falsifying
samples of 1000 consecutive points for normality, by testing simultaneously for skewness and excess
kurtosis, 65\% of the samples are rejected at a $p=0.01$ acceptance level. Bootstrapping
the same number of samples of 1000 points yield a rejection rate of 85\%. A Gaussian fit
to the noise histogram overestimates the average and the number of small outliers, and it
consistently underestimates the number of large outliers (Fig.\,\ref{fig:noise_properties}). The skewnormal distribution \citep[e.g.,]{azzalini1999}
\begin{equation}N_s(\xi,\omega,\alpha)=\frac{1}{\sqrt{2\pi\omega}} \exp\left(-\frac{(x-\xi)^2}{2\omega^2}\right) \left(1+\mbox{erf}\left[\alpha\frac{x-\xi}{\sqrt{2}\omega}\right]\right),\label{eq:skewnormal_distribution}\end{equation}
 is more appropriate to describe the overall noise specifications. For the CoRoT data,
we derive values of $\xi=0.14$ and $\omega=0.03$ for the location and scale parameters,
and a value of $\alpha=1.18$, which determines the shape of the distribution ($\alpha=0$ means the distribution
is normal, $\alpha>0$ means the distribution is right-skewed). These values imply right-skewed, leptokurtic distributed noise, with
a skewness $g_1\approx0.2$ and an excess kurtosis $g_2\approx0.1$.

Next, we simulate the influence of skewnormality on the parameter and error estimation of a model

\[F(t_i) = \mu + A\sin[2\pi(ft_i+\phi)].\]

To do so, we generate three collections of light curves, each set containing at least 500 light curves with highly skewed
noise ($\alpha=10$). To the first set of light curves, we add a high S/N monoperiodic sinusoidal signal (S/N$\sim 180$), to
the second set a low S/N monoperiodic sinusoidal signal (S/N$\sim 4$), and to the third set a superposition of
200 sinusoids with S/N between 3 and 190 (which mimics the CoRoT data of HD\,180642). To estimate $f$, we use the peak frequency in the Scargle periodogram of each light curve. The other
parameters are determined via ordinary linear regression. Subsequently, the distribution of each estimator is compared with the theoretical
formulae, as described by \citet{montgomery1999} but additionally taking correlation effects into account \citep{schwarzenberg2003}, e.g. the error estimate on the
frequency is
\begin{equation}\hat{\sigma}(f) = \sqrt{\frac{6}{N_{\rm eff}}} \frac{1}{\pi T} \frac{\sigma_r(t_i)}{a},\label{eq:frequency_error}\end{equation}
where $\sigma_r(t_i)$ is the standard deviation of the residuals. The effective number of observations $N_{\rm eff}$ is estimated by
counting the average distance between sign changes in the residuals. From Table\,\ref{tbl:scargle_comp_th_emp},
we can see that there is no discrepancy between an estimator and the real input value in the case of a monoperiodic, high S/N signal, besides
the fact that the `theoretical' error estimates, such as Eq. (\ref{eq:frequency_error}), are rather conservative. For a multiperiodic signal with low S/N, the same pattern emerges except
for the amplitudes: the estimator of the amplitude is slightly biased towards smaller values, but is still well within the error bars. The opposite bias is found in the signals with a low S/N value, but here an extra bias is introduced because peaks disappear in the noise for
low amplitude values.

Finally, we analyze the influence of skewness on the Scargle periodogram in a qualitatively way using a large number of simulations
of skew normally distributed noise with different parameters. We only find some additional noise at low frequencies, but this effect
is only apparent for very high $\alpha$ values. A set of heteroscedastic skew normal samples also introduces additional noise at low
frequencies, but again, the degree of heteroscedasticity has to be unrealistically high compared to the case of the CoRoT data, to have a significant influence.

In conclusion, although the deviation from uncorrelated homoscedastic white Gaussian noise is significant for the CoRoT data, it is not
 dramatic: we are dealing with a slightly right-skewed, leptokurtic distribution.
The above simulations suggest that a significant influence on parameter and error estimation is only noticable for high departures of normality.
Moreover, as will become clear in the following sections, the noise level is inherently low compared to the analyzed signals, and we are conservative in our significance criteria: in the following,
we adopt a $p$ value of $p=0.001$ in hypothesis testing, so that an order-of-magnitude estimate of $p$ is important, rather than a precise value. If at all,
only the correlation effects are worth taking into account for our analysis; this is done with the method outlined in \citet{schwarzenberg2003}.

\begin{table*}
\caption{Comparison between errors derived using theoretical formulas with correlation correction (input), and
empirically derived estimates of the parameters and errors (estimators), based on the results of more than 500 light curves of each 50\,000 simulated data points. The empirically derived value for the parameters are calculated as the average outcome of the simulations, while the error is determined as the standard deviation.\label{tbl:scargle_comp_th_emp}}
\centering\begin{tabular}{llllllll}\hline\hline
Set                    & Type & Frequency $f$                & $\sigma(f)$       & Amplitude $A$             & $\sigma(A)$           & Phase $\phi$                 & $\sigma(\phi)$\\\hline
Monoperiodic High S/N  &estimator& $5.4800000\pm 0.0000004$  & $0.000007$        & $2.0000 \pm 0.0002$       & $0.004$               & $+0.31703 \pm 0.00003$       & $0.0006$            \\
                       &    input& $5.4800000$               & $0.000007$        & $2.0000$                  & $0.004$               & $+0.31700$                   & $0.002$             \\
Monoperiodic Low S/N   &estimator& $5.47998  \pm 0.00002$    & $0.0008 $         & $0.2153 \pm 0.0001$       & $0.003$               & $+0.319   \pm 0.002$         & $0.06$              \\
                       &    input& $5.48000$                 & $0.0007$          & $0.0200$                  & $0.004$               & $+0.317$                     & $0.2$               \\
Multiperiodic          &estimator& $5.4868899\pm 0.0000003$  & $0.000007$        & $34912  \pm 3$            & $66$                  & $-0.03552\pm 0.00003$        & $0.0006$            \\
                       &    input& $5.4868900$               & $0.00001$         & $34918$                   & $98$                  & $-0.03551$                   & $0.003$             \\\hline
\end{tabular}

\end{table*}



\section{Modelling of the CoRoT light curve}

Most of the calculations concerning stellar oscillations of $\beta\,$Cep stars
assume modes with small amplitudes, to be able to treat multiperiodicity as a
linear superposition of multiple modes with infinite lifetime.  Fitting simple
sine functions, each with constant frequency, amplitude and phase through data
represents the simplest first order deviations from a theoretical equilibrium
state of the star.  However, when the perturbations are not confined to the
linear regime, higher order effects can only be modelled when different sines
are combined and/or harmonics are allowed for, spreading the signature of a
nonlinear effect in a Fourier periodogram over a wide range of frequencies.

Several physical origins of nonlinear effects in a light curve are plausible.
These include a nonlinear response of the stellar flux, leading to a distortion
of the light curve \citep[e.g.]{garrido1996}, nonlinear mode coupling through
resonant interaction between different modes \citep[e.g.]{dziem1982,
buchler1997}, excitation of strange-mode oscillations in highly nonadiabatic
regimes \citep[e.g.]{saio1998, glatzel1994}.  etc.  In particular, nonlinear
resonant mode coupling can be distinguished from complicated beating among
linear modes by checking the occurrence of frequency and/or phase locking, which
is not expected for a superposition of linear modes. Nonlinear oscillation
signatures may also include time-variable amplitudes or phases.

Given that we are dealing with the light curve of a large-amplitude $\beta\,$Cep
star, which is of unprecedented quality and quantity, it is not a priori clear
if a linear superposition of mode frequencies is the best approach to treat the
variability in the CoRoT light curve of HD\,180642. Therefore, we first perform
a traditional linear analysis of the light curve. Next, some nonlinear models
are constructed and fitted to the data, as well as compared with the fit
assuming linear mode frequencies. This comparison is done by means of
statistical criteria taking into account the number of free parameters. We thus
deduce the most likely physical interpretation of the variability of HD\,180642
from the data point of view.

\subsection{Superposition of linear modes}\label{sect:linear_analysis}

The first analysis of the CoRoT light curve of HD\,180642 was done according to
the traditional method, using the linear Scargle periodogram \citep{scargle1982}
and consecutive prewhitening, translating to a well-known model of the form
\begin{equation}F_1(t_i) = c + \sum_{j=1}^{n_f} A_j\sin[2\pi(f_jt_i+\phi_j)]
  \label{eq:model1}
\end{equation}
for $n_f$ frequencies. Here, $A_j$, $f_j$ and $\phi_j$ denote the amplitudes,
frequencies and phases. The model was evaluated at every time of observation
$t_i$. At each prewhitening stage, all amplitudes, phases and the constant
factor were refitted using the original light curve. This method implies a
frequency resolution of the order of the Rayleigh limit $1/T=0.0064$~d$^{-1}$,
making the frequency determination less precise when several frequencies are
confined to a region of this width. 

It is well known that nonlinear least squares fitting in the time domain while
leaving the frequencies, amplitudes and phases free, can improve seriously the
fit quality compared to the case where the frequency values are fixed to those
resulting from the periodogram, but also that the success of such a procedure is
largely dependent on the appropriate choice of good starting values, particularly
when many frequencies are present.  The starting values we adopted for the
amplitudes and phases are those that resulted from ordinary least squares
regression, while we fixed the frequencies from the Scargle periodogram.

\begin{figure}
\includegraphics[width=\columnwidth]{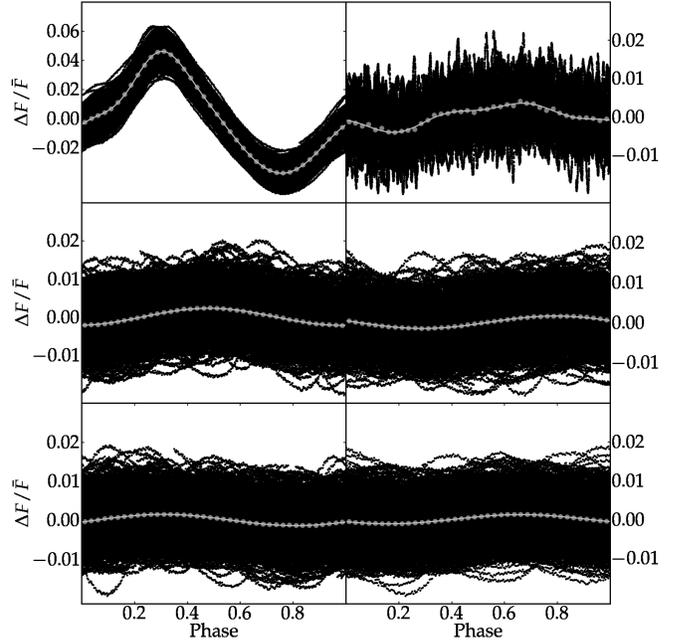}
\caption{Phase diagrams of first three independent frequencies. Grey lines are
fits, grey circles are averages of phase bins, black circles are data points
corresponding to successive prewhitening stages: (\emph{top left}) five
harmonics of $f_1=5.4868(9)$~d$^{-1}$, (\emph{top right}) three harmonics of
$f_2=0.2991(7)$~d$^{-1}$, (\emph{middle left to bottom right}) one harmonic of
$f_3=6.324(6)$~d$^{-1}$, $f_4=8.409(2)$~d$^{-1}$,$f_5=7.254(7)6)$~d$^{-1}$,
$f_6=11.811(6)$~d$^{-1}$.}\label{fig:phases_f1_f2}
\end{figure}

It was immediately clear from the first few detected frequencies that the
Scargle periodogram is not the optimal choice to describe or detect the
variability of HD\,180642, although it is most certainly a powerful indicator
and intuitive. At least three harmonics of the main frequency were detected,
with two more being marginally significant.  Also the second independent
frequency was best modeled with several harmonics (see upper right panel of
Fig.\,\ref{fig:phases_f1_f2}).  The amplitudes of the remaining frequencies are
small enough to be modelled by single sines, as illustrated for four of them in
Fig.\,\ref{fig:phases_f1_f2}.  This figure also shows that the light curve
cannot be adequately modelled by only a few frequencies and their harmonics.

\begin{figure*}
\includegraphics[width=2\columnwidth]{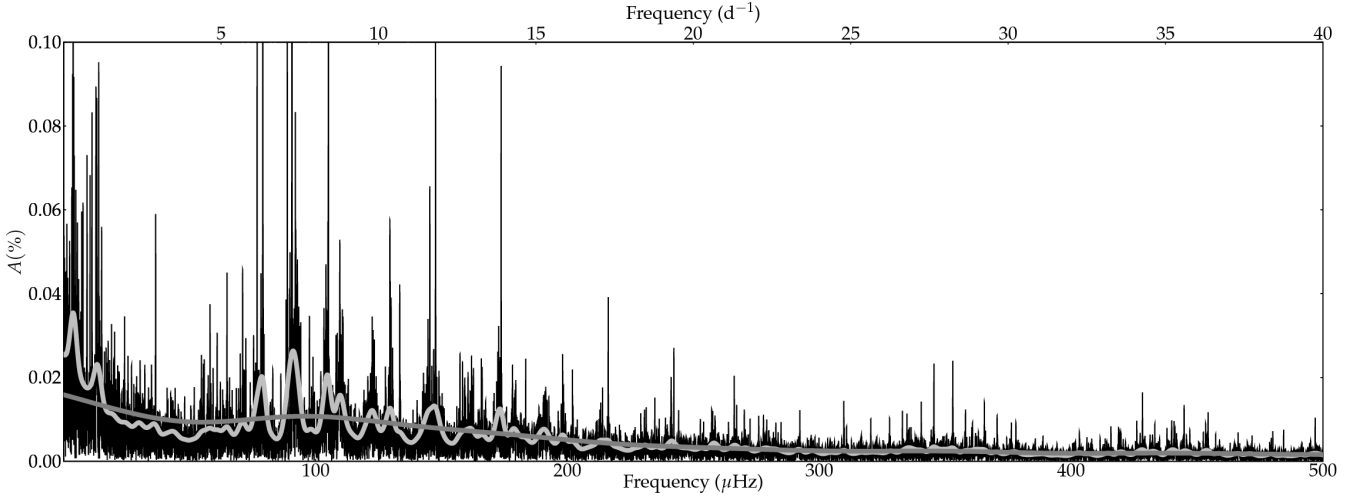}
\caption{Scargle periodogram (black) after prewhitening of dominant mode and its
harmonics. Overlays are Gaussian convolutions with $\sigma=0.1$~d$^{-1}$ (light
gray) and $\sigma=2$~d$^{-1}$ (dark gray), which are good indicators for power
excess and the empirical average noise level, respectively. The empirical noise
level only coincides with the theoretical noise level for white Gaussian Noise
at very high frequencies.}\label{fig:first_periodogram}
\end{figure*}

The result of this traditional analysis is a wealth in frequencies, excluding
clearly a monoperiodic model. We calculated up to 200 statistically significant
frequencies (Table\,\ref{raw_scargle}), although it has to be noted that
``only'' about 100 of them would be considered as not due to noise using the
traditional signal-to-noise criterion of \citet{breger1993}. However, there is
no doubt that the peaks are not due to noise because of two reasons. First, even
after prewhitening 200 frequencies with a nonlinear least squares fit, the residual amplitudes
are far above the instrumental noise level (discussed in Sect.\,\ref{sect:noise}),
which would be expected if the signal would be composed of a superposition of linear modes. This can also be seen in Fig.\,\ref{residues}, which we will discuss further in the text. Second, it is instructive to
describe the distribution of frequencies across the spectrum, to see where all
the frequencies reside. If most of the detected peaks are due to noise, they
should be more or less randomly distributed across the analyzed frequency
spectrum. In order to make the interpretation more clear, we decided to
prewhiten a model of the first dominant mode and its five significant harmonics,
or
\begin{equation}
F(t_i) = c + \sum_{j=1}^5 a_j\sin[2\pi(jf_1t_i + \phi_j)],
\label{eq:dominant_mode_first_model}
\end{equation}
where the initial harmonic fit was improved with a nonlinear least squares fit,
letting also the frequency variable but with fixed harmonic combinations. Then,
a power spectrum normalized by the total variance of the prewhitened data
(denoted as $Z$) was calculated.  This means that the expected noise level under
the assumption of Gaussian white noise corresponds to $Z=1$. The small deviation
from this assumption (see Sect.\,\ref{sect:noise}), implies that $Z=1$ slightly
underestimates the true noise level. Next, the
periodogram was averaged using Gaussian filters with $\sigma_1=0.1$~d$^{-1}$ (to
smooth out the peaks) and $\sigma_2=2$~d$^{-1}$ to estimate the empirical noise
level. The result is shown in Fig.\,\ref{fig:first_periodogram}.  Noticeable
power excess exists around $0.3$~d$^{-1}$, $1.0$~d$^{-1}$, $6.3$~d$^{-1}$,
$7.3$~d$^{-1}$, $8.4$~d$^{-1}$, $8.8$~d$^{-1}$,$9.8$~d$^{-1}$, $10.4$~d$^{-1}$,
$11.0$~d$^{-1}$, $12.3$~d$^{-1}$,$13.9$~d$^{-1}$, and, finally, to a lesser
extent also $14.15$~d$^{-1}$. Most of these power excess regions are not due to
one large peak, but represent a smoothing of many closely spaced peaks in the
periodogram, e.g., in the low frequency region ($\gtrsim 1$~d$^{-1}$). It is
clear that the low amplitude frequencies are \emph{not} due to noise, but is
actual signal; it is apparent that there are almost no frequencies nor any sign
of power excess in the region between $1$~d$^{-1}$ and $5$~d$^{-1}$. The higher
frequency regions ($>14$~d$^{-1}$) are much closer to the theoretical noise
level, but are at the same time contaminated by secondary window peaks.

As it turned out, several of the frequencies are linear combinations of 
other frequencies (Table \ref{tbl:combinations}). The influence of a few
frequencies is thus widely spread over the entire frequency spectrum. This is
why we consider the second model discussed below.

\subsection{Nonlinear frequency locking}

Frequencies excited through nonlinear resonant mode coupling can manifest
themselves in a natural way through combination frequencies, which may seem, at
first sight, independent from the others.  Such frequency locking is one effect
that can be derived from the Amplitude Equation formalism
\citep[e.g.]{dziem1982,buchler1997,vanhoolst1998}, if amplitudes and phases are
constant in time. Following this assumption, a summary of the most obvious
combination frequencies is given in Table\,\ref{tbl:combinations}. All
combination frequencies were identified following the method described in
\citet{degroote2009}. We selected only those combinations where the difference
between the true combination value and the real value is below half of the
Rayleigh limit $L_R=0.0064$\,d$^{-1}$. The nonlinear leakage can then be viewed
to spread over a wide range of the frequency spectrum, roughly between
$0$\,d$^{-1}$ and $20$\,d$^{-1}$.

This phenomenon of combination frequencies has been detected previously in
$\beta$ Cep stars, e.g.\ in $\nu\,$Eri \citep{handler2004} and in 12\,Lac
\citep{handler2006}. For these stars, only positive combinations were detected.
It was difficult, therefore, to interpret these combinations, either in terms of
light curve distortions due to nonlinear response or due to nonlinear resonant
mode coupling. Indeed, both these phenomena would naturally give rise to
difference combination frequencies as well as phase locking, which were not
detected in these two stars.  Moreover, third order combinations were not
unambiguously identified, either because the amplitudes were too low, or they
were not excited.

Under the assumption that the combination frequencies are real in HD\,180642,
the amplitudes, phases and independent frequencies were refitted using the
previous values as starting values, while fixing the dependent frequencies
according to their linear combination throughout the fit:
\begin{equation}
\begin{array}{ll}
\displaystyle F_2(t_i) = c 
&\displaystyle + \sum_{k=1}^{n_f}A_k\sin[2\pi(f_kt_i+\phi_k)]\\
\displaystyle &\displaystyle + \sum_{l=1}^{m_f}A_l\sin[2\pi(f_lt_i + \phi_l)]\\
\end{array}\label{eq:model2}
\end{equation}
with
\[f_l = n_l^1f_l^1 + n_l^2f_l^2,\]
a linear combination of two independent frequencies. It is assumed that there
are $n_f$ independent frequencies, and $m_f$ dependent frequencies.

\begin{figure}
\includegraphics[width=\columnwidth]{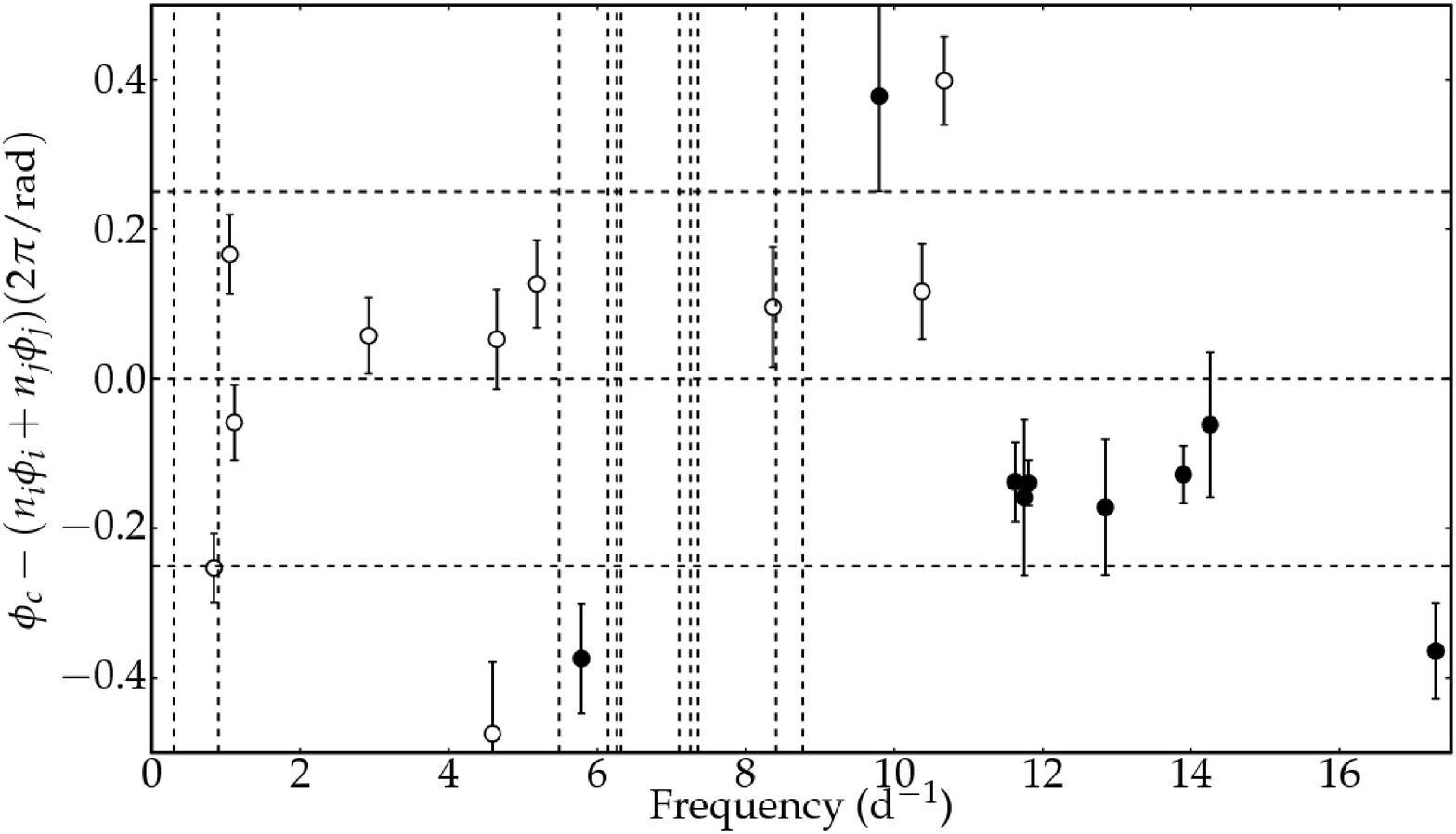}         
\includegraphics[width=\columnwidth]{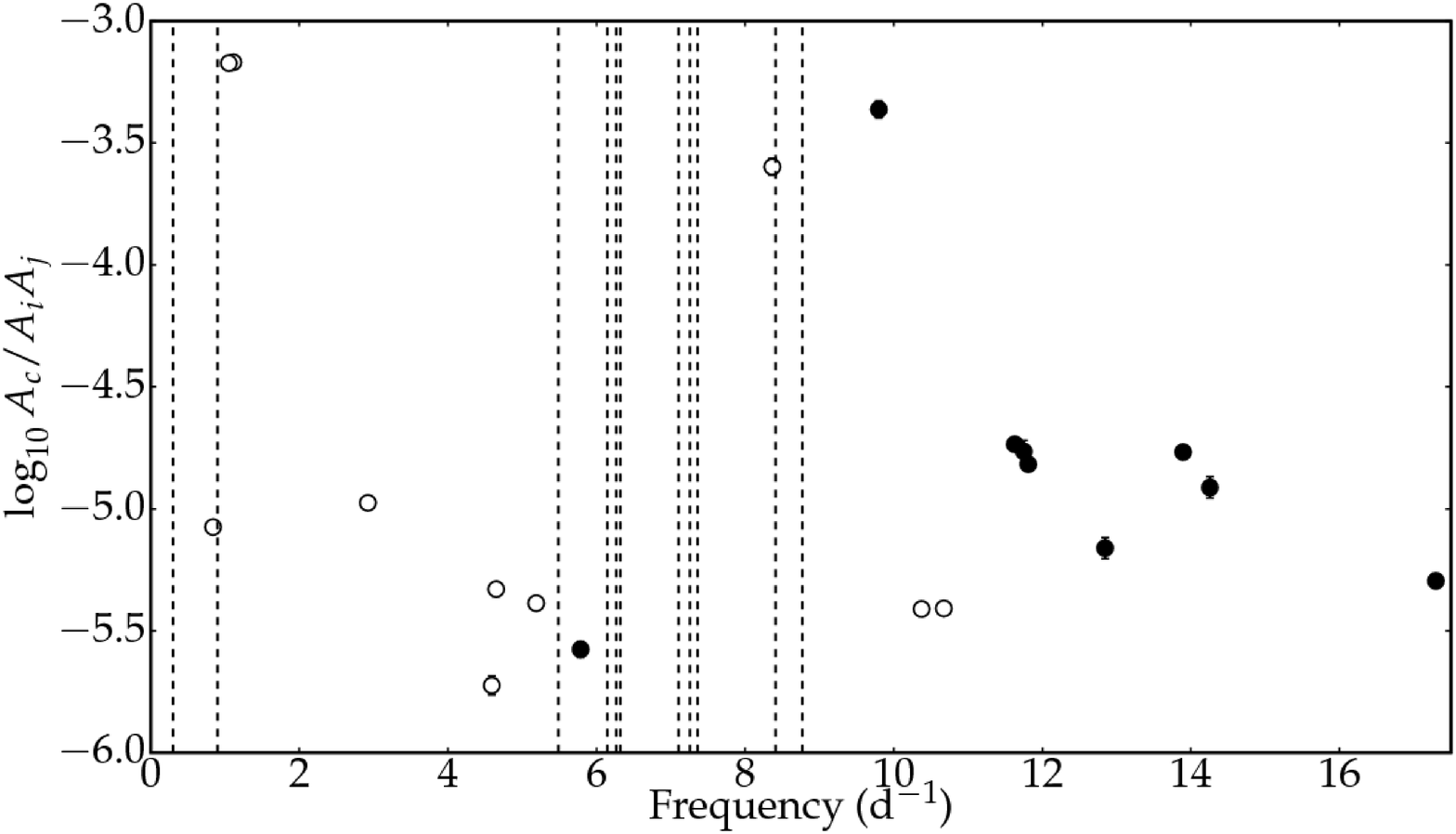}
\includegraphics[width=\columnwidth]{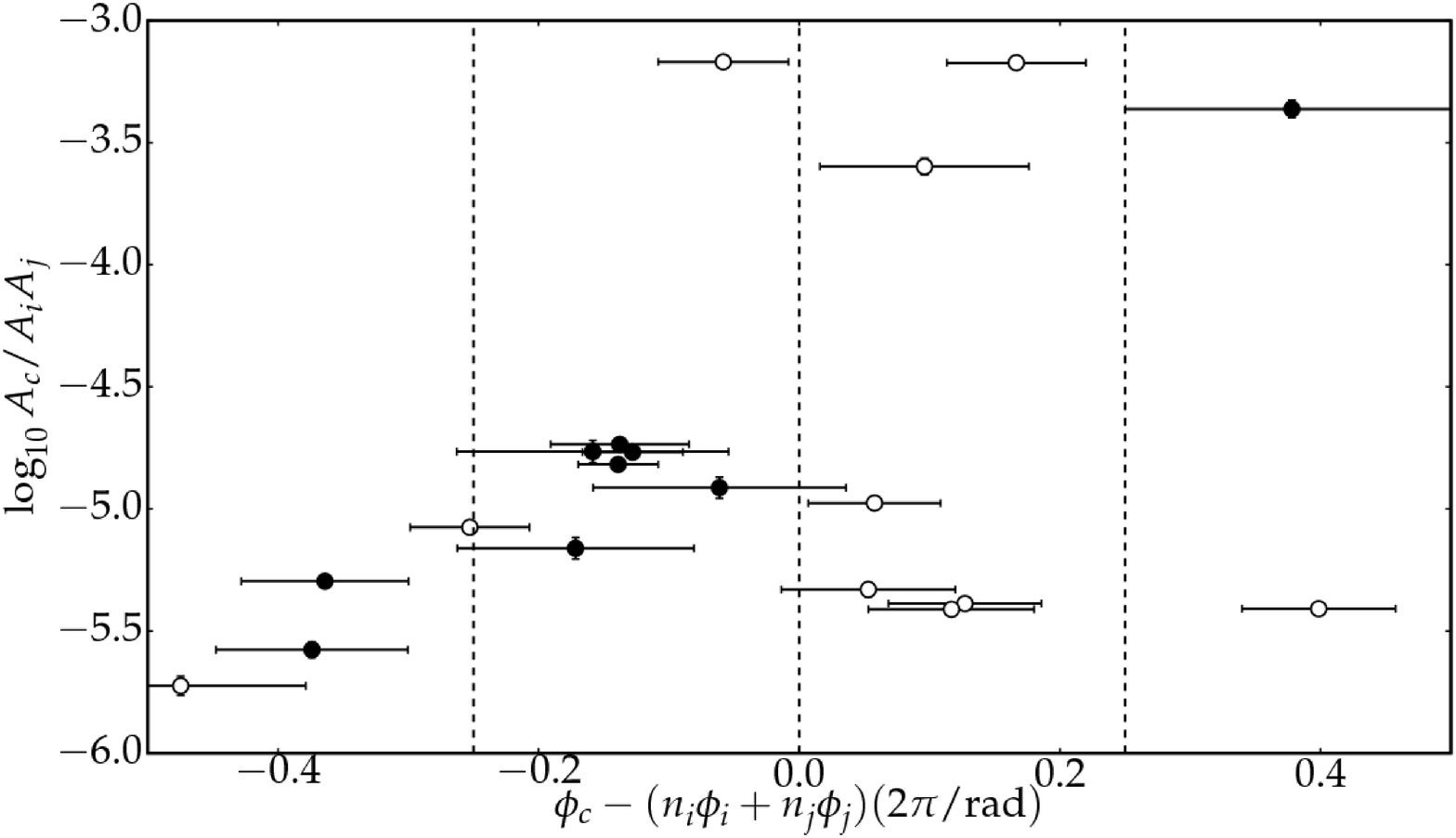}
\caption{(\emph{top}) Relative phase as a function of frequency. Error bars
denote $3\,\sigma$ level. The parent modes are indicated by vertical
lines. Horizontal lines denote a $\pi/2$ phase lag and lead with respect to the parent
modes. Sum combinations are filled circles, difference combinations are open
circles. (\emph{middle}) Relative amplitudes as a function of
frequency. (\emph{bottom}) Relative amplitude as a function of
relative phase.}\label{fig:combinations}
\end{figure}

In order to better describe the combination frequencies and their origin, their
relative phases and amplitudes were analysed. Following \citet{buchler1997}
and \citet{vuille2000a}, they are defined as
\[\phi_r=\phi_c-[n_i\phi_i+n_j\phi_j]\]
and
\[A_r = \frac{A_c}{A_iA_j}\]
with the subindex $i$ referring to the parent mode with the largest amplitude,
$j$ to the parent mode with the smallest amplitude and the index $c$ to the
daughter mode. 
The relative phases and amplitudes for all candidate combination
frequencies, except harmonics, from Table\,\ref{tbl:combinations} are shown in
Fig.\,\ref{fig:combinations}. 

\begin{figure*}
\includegraphics[width=2\columnwidth]{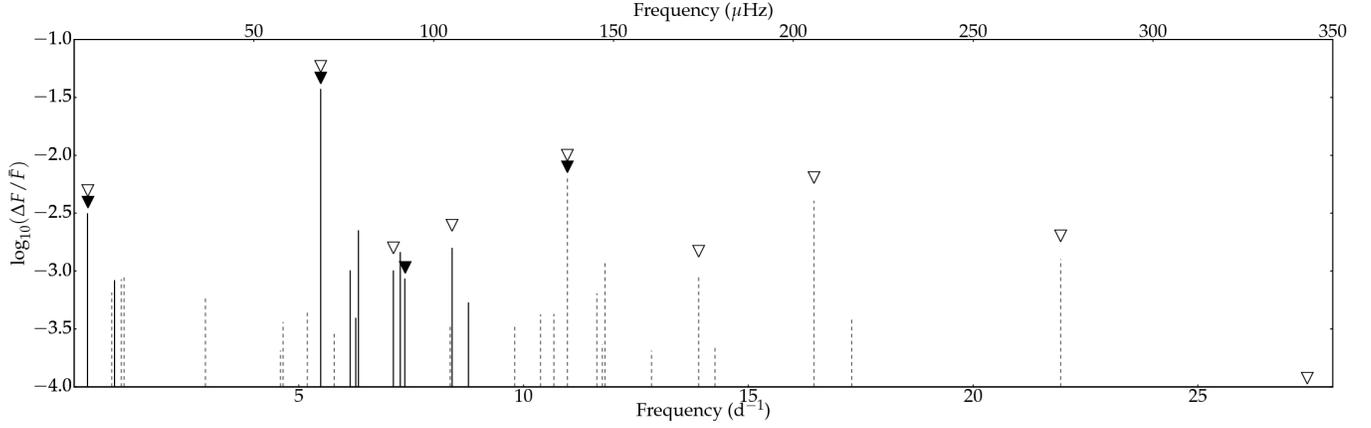}
\caption{Summary of independent (full lines) and combination frequencies (dashed
  lines) for the nonlinear frequency locking model $F_2$ described in the
  text. The frequencies detected in the ground-based photometry and spectroscopy by \citet{briquet2009} are indicated by closed and open triangles, respectively.}
\label{bars}
\end{figure*}

To make the discussion more readable, we denote
each daughter mode by a unique designation,
\[d_{i,j}(n_i,n_j)\]
where $i$ and $j$ are indices of the largest and smallest amplitude parent modes
respectively (the higher this index, the lower the amplitude), and $n_i,n_j$ are
the corresponding coefficients in the linear combinations
(Table\,\ref{tbl:combinations}). Sum frequencies are distinguished from
differences by the sign of the coefficients.
Several interesting features appear:
\begin{enumerate}
\item There are six daughter frequencies which have four properties in common:
they are a sum of the dominant mode with another frequency, they cluster around
the same relative phase ($\sim 0.15$), they have comparable relative amplitudes
and they have the same first-order coefficients ($n_i=n_j=1$): $d_{1,3}(1,1)$,
$d_{1,4}(1,1)$, $d_{1,6}(1,1)$, $d_{1,8}(1,1)$, $d_{1,10}(1,1)$ and
$d_{1,11}(1,1)$.
\item A similar clustering around a common relative phase is visible for 6
differences: $d_{4,8}(1,-1)$, $d_{1,4}(-1,1)$, $d_{1,3}(2,-1)$, $d_{1,2}(1,-1)$,
$d_{5,6}(2,-1)$ and $d_{1,2}(2,-2)$, although they have different coefficients
and different relative amplitudes.
\item Around the second harmonic of the dominant mode, a spacing with $\Delta
F=0.29917$\,d$^{-1}$ is clearly visible. The daughter frequencies
$d_{1,2}(2,-1)$, $d_{1,2}(2,-2)$ have almost the same relative amplitudes, and
are both found in two consecutive prewhitening stages.
\item The daughter frequency $d_{1,3}(1-1)$ has a relative phase of $\pi/2$
\end{enumerate}
The result for the fit using model $F_2$ is provided in
Table\,\ref{tbl:fixed_combos} while a summary diagram of the independent and
combination frequencies is shown in Fig.\,\ref{bars}.

\begin{table}
\centering 
\caption{Parents ($p_1,p_2$) and their orders ($n_1,n_2$) of 
combination frequencies $f_c$. The Rayleigh limit is equal to
$L_R=0.0064$~d$^{-1}$. Frequency values $f_1,f_2$ and $f_c$ correspond to the
highest peaks in successive prewhitening stages. $\Delta=|n_1f_1 + n_2f_2 -
f_c|$ denotes difference between the true linear combination and the found
value. Column ID shows the unique designation of daughter mode $f_c$ (see
text).}\label{tbl:combinations}
\begin{tabular}{lrrrrrl}\hline\hline
ID & $n_1$ & $f_1$ (d$^{-1}$) & $n_2$ & $f_2$ (d$^{-1}$) & $f_c$ (d$^{-1}$) & 
$\Delta$ (d$^{-1}$)\\\hline
$d_{1,3}(1,1)$  & 1 & 5.48689 & 1  & 6.32482 &  11.81164 & 0.00006\\
$d_{1,3}(-1,1)$ &-1 & 5.48689 & 1  & 6.32482 &   0.83794 & 0.00001\\
$d_{1,3}(2,1)$  & 2 & 5.48689 & 1  & 6.32482 &  17.29841 & 0.0002 \\
$d_{1,3}(2,-1)$ & 2 & 5.48689 & -1 & 6.32482 &   4.64845 & 0.0005 \\
$d_{1,2}(1,1)$ & 1 & 5.48689 &  1 & 0.29917 &   5.78662 & 0.0006 \\
$d_{1,2}(1,-1)$ & 1 & 5.48689 & -1 & 0.29917 &   5.18781 & 0.00009 \\
$d_{1,2}(2,-1)$ & 2 & 5.48689 & -1 & 0.29917 &  10.67458 & 0.00002 \\
$d_{1,2}(2,-2)$ & 2 & 5.48689 & -2 & 0.29917 &  10.37493 & 0.0005 \\
$d_{1,2}(1,-3)$ & 1 & 5.48689 & -3 & 0.29917 &   4.58920 & 0.0002 \\
$d_{1,4}(1,1)$  & 1 & 5.48689 &  1 & 8.40918 &  13.89585 & 0.0002 \\
$d_{1,4}(-1,1)$  &-1 & 5.48689 &  1 & 8.40918 &   2.92159 & 0.0007 \\
$d_{1,6}(1,1)$  & 1 & 5.48689 &  1 & 6.14336 &  11.63039 & 0.0001 \\
$d_{1,8}(1,1)$  & 1 & 5.48689 &  1 & 7.35867 &  12.84432 & 0.001 \\
$d_{1,10}(1,1)$ & 1 & 5.48689 &  1 & 8.77086 &  14.25740 & 0.0003 \\
$d_{1,11}(1,1)$ & 1 & 5.48689 &  1 & 6.26517 &  11.75173 & 0.0003 \\
$d_{5,6}(1,-1)$ & 1 & 7.25476 & -1 & 6.14336 & 1.11216 & 0.0008 \\
$d_{5,6}(2,-1)$ & 2 & 7.25476 & -1 & 6.14336 & 8.36870 & 0.0025 \\
$d_{4,8}(1,-1)$ & 1 & 8.40918 & -1 & 7.35866 & 1.04985 & 0.0007 \\
$d_{7,9}(1,3)$ & 1 & 7.10353 &  3 & 0.89870 & 9.79999 & 0.0003 \\
\hline\hline
\end{tabular}
\end{table}

In the case of HD\,180642, we hence do see difference combination frequencies,
in contrast to the cases of the two large-amplitude $\beta\,$Cep stars
$\nu$\,Eri \citep{handler2004} and 12\,Lac \citep{handler2006}.  These difference
frequencies are still well above 0.1\,d$^{-1}$, and are thus in the regime of white noise (see Fig.\ref{fig:noise_properties}) Such low
combination frequencies are expected to occur with similar amplitudes as the
sum combinations, for both the nonlinear distortion model and a nonlinear
resonant mode coupling model. While we see more sum frequencies than
differences, we do reach the regime of g-mode frequencies through several
combinations for HD\,180642. We also found higher order combinations here, up to
order four (see Table\,\ref{tbl:combinations}), than for $\nu\,$Eri and
12\,Lac. 

We note from Fig.\,\ref{fig:combinations} that four combination frequencies have
a much higher $A_r$-value than the others. This is simply due to the fact that
these are the four combinations not involving the dominant mode (hence the
denominator in the definition of $A_r$ is much smaller).  Further, the relative
amplitudes of the combinations involving the dominant mode cover a range of a
factor ten and the phases cover the entire range $[-0.5,0.5]$, although several
relative phases of difference frequencies are equal within the error bars, and
similarly for the sum frequencies. The largest relative amplitudes all occur for
a sum frequency due to a three-mode resonance model involving the dominant
mode. The difference frequencies of the same three modes, if they occur, all
have lower relative amplitude. We interpret this as due to nonlinear resonant
mode locking, as a nonlinear distortion would not privilege larger amplitudes
for sum or difference frequencies.  

In principle, the relative amplitudes of the resonantly locked frequencies can
help to constrain the mode degrees, because the geometric cancelling effect is
different for different degrees. Unfortunately, we cannot use the relative
amplitude values to derive the mode degrees, because all large-amplitude three
mode resonances involve the dominant radial mode which does not imply geometric
cancelling, and we have no other information on the degrees of the parent
frequencies. 

\subsection{Time dependent amplitudes and phases}

In the previous section, we assumed the amplitudes and phases to be constant in
time. However, the Amplitude Equation formalism also allows for solutions where
this is not the case \citep{buchler1997}. Amplitude and phase modulations may
occur, which can be (multi)periodic or chaotic. The light curve of HD\,180642 as
measured with CoRoT is of such high quality, that it becomes possible to detect
and model these variations through changes in the highly sampled phase
profile. Although a sine function with five harmonics is a good fit in the phase
diagram (Fig.\,\ref{fig:phases_f1_f2}), it is also clear from the same figure
that this model is only an `average' model; in fact, the fit is not optimal to
model a particular phase. Some phases can be modelled adequately with three
harmonics, others need four, etc. Moreover, the minima and maxima seem to
oscillate around an equilibrium value.

To quantify this time-dependent behaviour, a harmonic fit was calculated for
every covered phase of the main frequency. The number of harmonics to be used is
determined from the $\chi^2$ statistic of the data with respect to the
model. The number of significant harmonics was taken as the lowest one that
achieves a $\chi^2<1.5$. This number varies mainly between three and four, with
few exceptions.

We quantify the complexity of each phase profile by the ratio of the
harmonic's amplitudes compared to the main amplitude. The higher this ratio, the
more significant the specific harmonic is. Also, from each fit, we extract the
fitted constant as an indicator for long term trends. Finally, peak-to-peak
variations in the phases are calculated. Using these methods, we finally arrive
at an adapted version of Eq.\,(\ref{eq:dominant_mode_first_model}):

\begin{equation}
F_3(t_i) = c(t_i) + \sum_{j=1}^5 a_j(t_i)\sin[2\pi(jf_1t_i + \phi_j(t_i))],
\label{eq:dominant_mode_time_variable}
\end{equation}
with
\[
\begin{array}{rcl}
\displaystyle c(t_i)       & = & \displaystyle C+\sum_{k} A^c_k\sin[2\pi(f^c_kt_i+\phi^c_k)], \\
\displaystyle a_j(t_i)     & = & \displaystyle A+\sum_{l} A^{a_j}_l\sin[2\pi(jf^{a_j}_lt_i+\phi^{a_j}_l)], \\
\displaystyle \phi_j(t_i)  & = & \displaystyle \Phi+\sum_{m} A^{\phi_j}_m\sin[2\pi(jf^{\phi_j}_mt_i+\phi^{\phi_j}_m)]\cdot\end{array}
\]

\begin{figure*}
\includegraphics[width=\columnwidth]{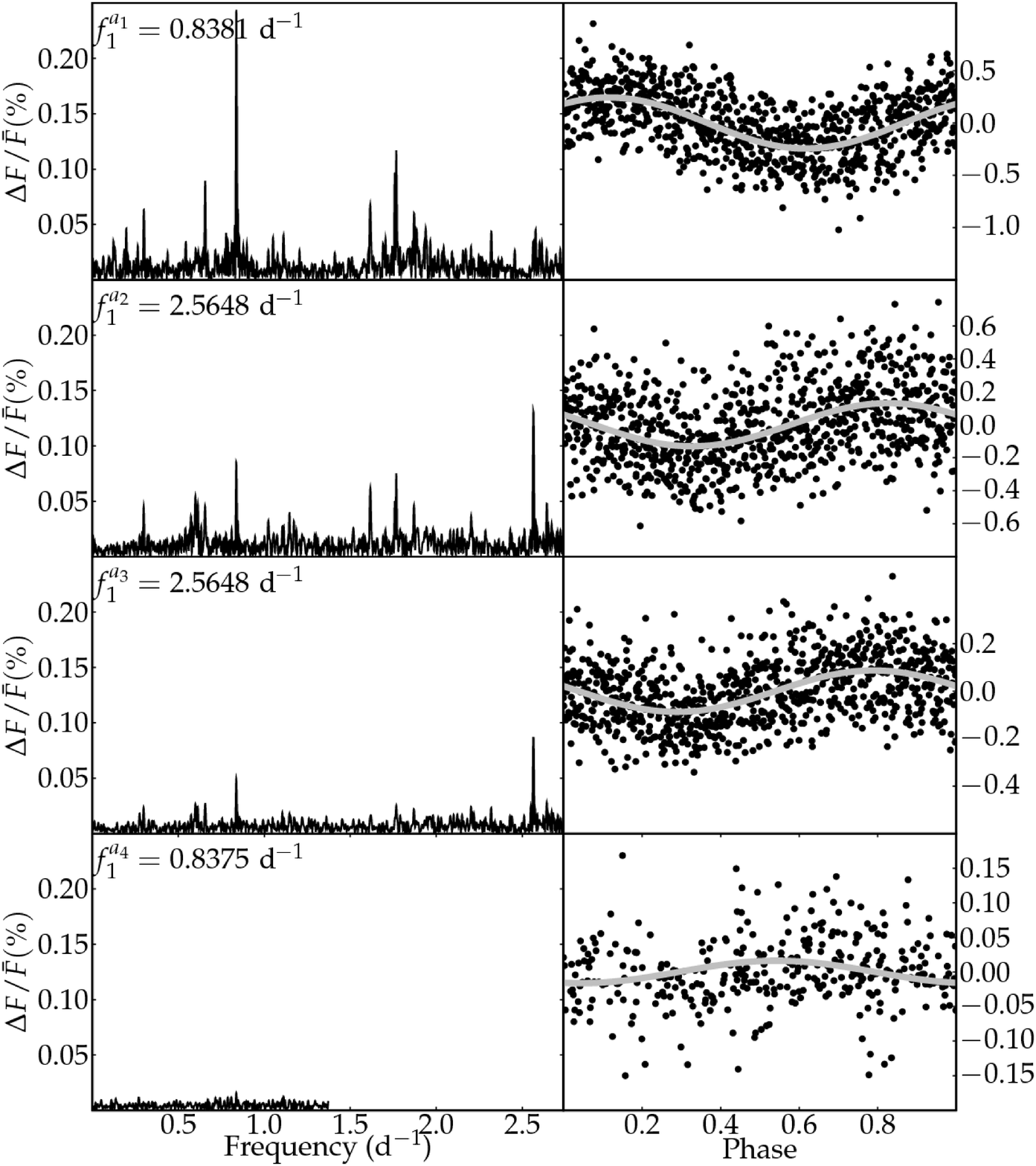}
\includegraphics[width=\columnwidth]{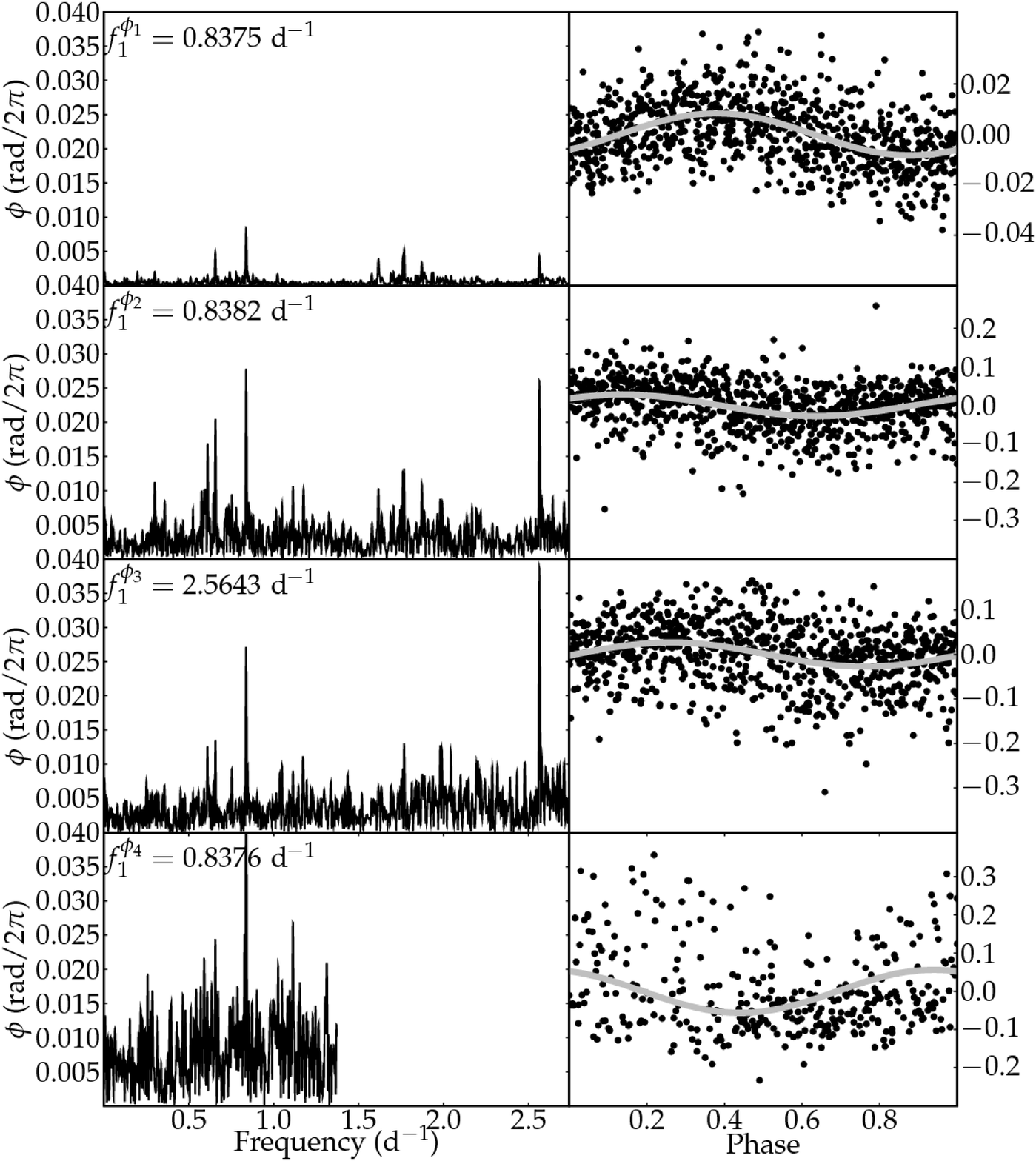}
\caption{Periodograms and phase diagrams for the time-variable amplitude
  ($f_1^{a_i}$) and phase ($f_1^{\phi_i}$) of the first four harmonics of the
  dominant mode.  Because the fourth harmonic was not detected in every phase,
  there are less points in the lower panels for which the Nyquist frequency is
  also lower than for the other harmonics.}\label{fig:variables}
\end{figure*}

For clarity, we first examine what the linear interpretation of this model would be. The simple model with $k=0$ and $j,l,m=1$
 can be linearized with the assumption that $A^\phi\leq1$. Violation of this assumption only influences the amplitude determination. Linearizing (\ref{eq:dominant_mode_time_variable}) gives
\begin{equation}
 \begin{split}
  F_3'&(t_i) = A\sin(Ft+\Phi)\\
         &+ A^a/2\sin[(F-f^a)t_i + (\Phi-\phi^a+\pi/2)] \\
         &+ A^a/2\sin[(F+f^a)t_i + (\Phi+\phi^a-\pi/2)] \\
         &+ AA^\phi/2\sin[(F+f^\phi)t_i + (\Phi+\phi^\phi)] \\
         &+ AA^\phi/2\sin[(F-f^\phi)t_i + (\Phi-\phi^\phi+\pi)] \\
         &+ A^aA^\phi/4\sin[(F+f^a-f^\phi)t_i + (\Phi+\phi^a-\phi^\phi+\pi/2)]\\
         &+ A^aA^\phi/4\sin[(F-f^a-f^\phi)t_i + (\Phi-\phi^a-\phi^\phi-\pi/2)]\\
         &+ A^aA^\phi/4\sin[(F+f^a+f^\phi)t_i + (\Phi+\phi^a+\phi^\phi-\pi/2)],
 \end{split}\label{eq:linear_expansion}
\end{equation}
or when $f^a=f^\phi$
\begin{equation}
 \begin{split}
F(t_i) &= A_{1}\sin(Ft + \Phi_{1})\\
       &+ A_{2}\sin[(F-f^a)t + \Phi_{2}]\\
       &+ A_{3}\sin[(F+f^a)t + \Phi_{3}]\\
 \end{split}\label{eq:linear_expansion_short}
\end{equation}
with, under the assumption that $A^\phi\ll A$,
\[A_{1}^2 = A^2 + \left(\frac{A^aA^\phi}{4}\right)^2 + \frac{AA^aA^\phi}{2}\cos(\phi^\phi-\phi^a-\pi/2)\approx A^2,\]
\[\Phi_1 = \arctan\left(\frac{A\sin\Phi+\frac{A^aA^\phi}{4}\sin(\Phi+\phi^a-\phi^\phi+\pi/2)}{A\cos\Phi+\frac{A^aA^\phi}{4}\cos(\Phi+\phi^a-\phi^\phi+\pi/2)}\right)\approx \Phi.\]
The results of the frequency analysis for the amplitudes, phases and constants
for this model assumption can be found in Tables\,\ref{tbl:var_3} to
\ref{tbl:var_10}. The analysis of the first frequency of the time-dependent
amplitude and phase for each harmonic is shown in Fig.\,\ref{fig:variables}.

With this last expansion in mind, we can use the results of the frequency
analysis of the amplitudes and phases to predict the occurrence of spurious
frequencies, actually originating from the linear expansion of the nonlinear
model, and thus not physically inherent to the star. We performed this exercise
for the primary component of the main radial mode and its harmonics. If we
assume that the primary component $f_1$ and its harmonics have a variable
amplitude and phase with frequency $f=0.8379$~d$^{-1}$, then we can already
explain the occurrence of several of the observed frequencies (see
Table\,\ref{tbl:nonlinear_predictions}). Obviously, not all predicted
frequencies are discovered. After fitting the light curve with this model and
inspecting the periodogram, we can see that there are additional secondary peaks
introduced which are not observed. When taking more frequencies into account, at
least some of the secondary peaks seem to cancel out. This motivated us to
construct several time-dependent amplitude models and to compare the fits. From
Tables\,\ref{tbl:var_3} to \ref{tbl:var_10}, we selected the most probable
frequencies and let the amplitudes and phases vary accordingly. First, we tried
with only one frequency. Gradually we added more frequencies, until we arrived
at a maximum of four frequencies determining the amplitude time variability.

The two frequencies 0.836\,d$^{-1}$ and 1.767\,d$^{-1}$ clearly stand out: the
amplitudes and phases oscillate on this time scale, but they are also found when
applying model $F_1$ given by Eq.\,(\ref{eq:model1}), which means that the
`entire' light curve is also oscillating at the same rate. This is confirmed by
analyzing the residuals after removing the constructed models: both of the
frequencies are recovered.

Despite the success of these time-dependent amplitude and phase 
models to explain several peaks in the periodogram,
we have to compare them more rigorously with the models of the forms
Eqs\,(\ref{eq:model1}) and (\ref{eq:model2}), which is the topic of the next
section. 

\begin{table}\caption{Selection of predicted versus observed frequencies under
    the assumption of time dependent amplitudes and phases 
(model given by Eq.\,(\ref{eq:dominant_mode_time_variable})).}
 \begin{tabular}{rrrl}\hline
component & secondary & spurious frequency & Error\\\hline $f_1$ & 0.8379 &
6.3248 & Observed\\ $f_1$ & -0.8379 & 4.6490 & Observed\\ $2f_1$ & 0.8379 &
11.8168 & Observed\\ $2f_1$ & -0.8379 & 10.1359 & -\\ $3f_1$ & 0.8379 & 17.2986
& Observed\\ $3f_1$ & -0.8379 & 10.1359 & -\\\hline
 \end{tabular}
 \begin{tabular}{rrrl}\hline
component & secondary & spurious frequency & Error\\\hline
$f_1$  &  1.7678 &  7.2546 & Observed\\
$f_1$  & -1.7678 &  3.7191 & -\\
$2f_1$ &  1.7678 & 12.7412 & Observed\\
$2f_1$ & -1.7678 &  9.2060 & -\\
$3f_1$ &  1.7678 & 18.2285 & -\\
$3f_1$ & -1.7678 & 14.6928 & -\\\hline
 \end{tabular}
 \begin{tabular}{rrrl}\hline
component & secondary & spurious frequency & Error\\\hline
$f_1$  &  2.5650 &  8.0519 & -\\
$f_1$  & -2.5650 &  2.9219 & Observed\\
$2f_1$ &  2.5650 & 13.5388 & -\\
$2f_1$ & -2.5650 &  8.4088 & Observed\\
$3f_1$ &  2.5650 & 19.0257 & -\\
$3f_1$ & -2.5650 & 13.8957 & Observed\\\hline
 \end{tabular}
\label{tbl:nonlinear_predictions}
\end{table}

\subsection{Model evaluation}

To compare the different models with each other and determine their
goodness-of-fit, we compute four evaluation statistics: the variance reduction
(VR), both Akaike's Information Criterion (AIC) and the Bayesian Information
Criterion (BIC) in the time domain, and the power reduction (PR) in the
frequency domain. The AIC is defined as
\begin{equation}
{\rm AIC} = 2k - 2\ln\mathcal{L_{\rm max}},
\label{eq:AIC}
\end{equation}
where $\mathcal{L_{\rm max}}$ is a maximum likelihood estimator (MLE), $n$ is
the effective number of observations and $k$ is the number of free parameters in
the model. Under the assumption of Gaussian white noise, we can insert the MLE
of the noise variance, $\hat{\sigma_i}^2={\rm RSS}/n$ with RSS the residual sum
of squares. Criterion (\ref{eq:AIC}) then becomes
\[{\rm AIC} = n\ln({\rm RSS}/n) + 2k + n .\]
The AIC can be calculated for every model, but is only relevant in comparisons;
the lower the AIC, the better the model. 

Despite the fact that the AIC
discourages the use of too many free parameters (unlike the variance and power
reduction), the BIC is more suited when we want to stress the importance of
simpler models over more complicated ones and thus increase the penalty for
introducing new parameters. The BIC is defined as
\begin{equation}
{\rm BIC} = -2 \ln\mathcal{L_{\rm max}} + k \ln n .
\label{eq:BIC}
\end{equation}
Analogous to the AIC, we can simplify this to
\[{\rm BIC} = n \ln \left({\rm RSS}/n\right) + k\ln n .\] 
The BIC is also only relevant in comparisons, where again the lower the BIC, the
better the model. We choose to use the BIC for our model selection rather than
the criteria, because we want to favour simple physically appropriate models
with the fewest possible degrees of freedom. The BIC is the most conservative
criterion in this respect, because it requires that the gain in variance
reduction must be worth the cost of introducing extra parameters.

\begin{figure}
\includegraphics[width=\columnwidth]{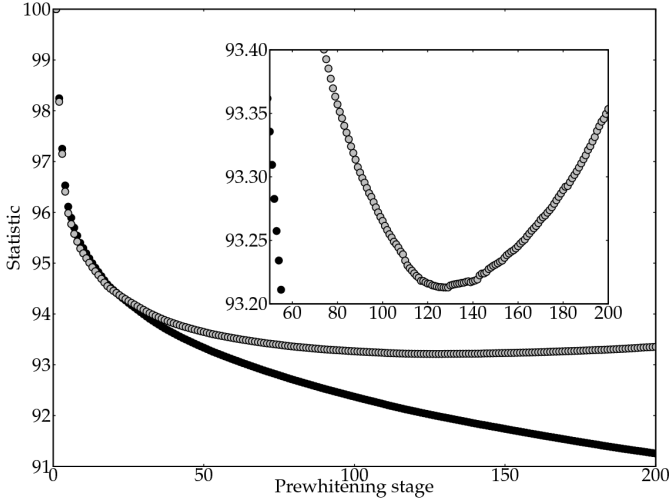}
\caption{Comparison of the AIC (black) and BIC (grey) for the Scargle analysis
of model $F_1$. The lower the value, the better the fit. The absolute values
have no meaning, so the AIC and BIC cannot be compared with each other. Clearly
the BIC gives a higher penalty for introducing extra parameters than the
AIC. From frequency 127 on, the BIC discourages the use of additional sines in
the model fit (inset is a zoom).}\label{fig:aicbic}
\end{figure}

First, we computed the AIC and BIC for all prewhitening stages. Indeed, given
the richness of the frequency spectrum, it could well be that not all
variability can be modeled \emph{adequately} with sine functions. Of course any
type of variability \emph{can} be considered as such, but this may result in the
use of too many parameters. As can be seen on Fig.\,\ref{fig:aicbic}, the BIC
sets the optimal number of sines to use for model fit $F_1$ to 127; the
introduction of 3 additional parameters is not worth the gain in variance
reduction any longer.

From Table\,\ref{tbl:model_evaluation}, we can see that {\it all statistics
prefer the model where the combination frequencies were fixed}, except for the
variance reduction. That the latter is slightly worse is no surprise, as the
number of free parameters is much smaller with fixed frequencies.

All versions of the model $F_3$ described by
Eq.\,(\ref{eq:dominant_mode_time_variable}) lead to a worse fit to the data than
similar models $F_1$ (in the sense of which peaks they can explain) of the form
of Eq.\,(\ref{eq:model1}).  The differences are small in most cases, but there
is a difference nonetheless. This was already foreseen in
Table\,\ref{tbl:nonlinear_predictions} because not all predicted peaks are
detectable, hence the fit introduces spurious peaks. This effect can also be
deduced from the PR: although more parameters, the fits explain the periodogram
less well than simpler models.

\begin{table}
 \caption{Model evaluation. The AIC and BIC criteria are compared to a model
 with only the first frequency prewhitened. The power reduction with respect to
 the original periodogram is computed by numerical integration of the model's
 amplitude periodogram. $N_p$ is the number of parameters used in the model. The
 model numbers between brackets refer to the Equations in the text. For each
 alternative model Eqs\, (\ref{eq:model2}) and
 (\ref{eq:dominant_mode_time_variable}) separated by horizontal lines, the
 closest linear (original) model Eq.\,(\ref{eq:model1}) is computed, and fitted
 through nonlinear least squares.  The best model according to the four
 considered statistics are emphasized in bold.}
\tabcolsep=4pt
\begin{tabular}{lcccccc}
\hline\hline
Model & AIC (\%) & BIC (\%)  & VR (\%) & PR (\%) &$N_p$\\\hline
Eq.\,(\ref{eq:model2}), fixed comb & \textbf{94.010} & \textbf{94.065} & 97.755          & \textbf{25.50} &  78\\
Eq.\,(\ref{eq:model1}), Scargle freqs       & 94.023          & 94.171          & 97.761          & 25.49          & 100\\
Eq.\,(\ref{eq:model1}), NLLS freqs  & 94.021          & 94.169          & \textbf{97.762} & 25.49          & 100\\ \hline
Eq.\,(\ref{eq:dominant_mode_time_variable}), 1 freq & 95.417          & 95.378          & 96.873          & 30.81 & 40\\
Eq.\,(\ref{eq:model1}), Scargle freqs                    & 95.405          & 95.314          & 96.872          & \textbf{30.87} & 28\\
Eq.\,(\ref{eq:model1}), NLLS freqs               & \textbf{95.404} & \textbf{95.314} & \textbf{96.872} & 30.85 & 28\\
\hline
Eq.\,(\ref{eq:dominant_mode_time_variable}), 2 freqs & 95.228 & 95.255 & 97.019 & 30.14 & 58\\
Eq.\,(\ref{eq:model1}), Scargle freqs                    & 95.204 & 95.130 & 97.017 & \textbf{30.24} & 34\\
Eq.\,(\ref{eq:model1}), NLLS freqs               & \textbf{95.204} & \textbf{95.130} & \textbf{97.017} & 30.22 & 34\\
\hline
Eq.\,(\ref{eq:dominant_mode_time_variable}), 3 freqs & 94.887          & 97.974 & 97.255 & 28.65 & 76\\
Eq.\,(\ref{eq:model1}), Scargle freqs                    & 94.837          & 94.783 & 97.263 & \textbf{28.84} & 43\\
Eq.\,(\ref{eq:model1}), NLLS freqs               & \textbf{94.837} & \textbf{94.783} & \textbf{97.263} & 28.82 & 43\\
\hline
Eq.\,(\ref{eq:dominant_mode_time_variable}), 4 freqs & 94.746 & 94.901 & 97.355 & 27.99 & 94\\
Eq.\,(\ref{eq:model1}), Scargle freqs                    & 94.687 & 94.664 & 97.361 & 28.29 & 52\\
Eq.\,(\ref{eq:model1}), NLLS freqs               & \textbf{94.687} & \textbf{94.664} & \textbf{97.361} & \textbf{28.29} & 52\\
\hline
\end{tabular}\label{tbl:model_evaluation}
\end{table}

\section{Time-frequency behaviour of parent and combination frequencies}

The enormous complexity of the power spectrum of HD\,180642 as measured with
CoRoT and many of the previous remarks raise questions about the stability of
the observed and treated modes in terms of amplitudes and frequencies.
Nonlinear resonant mode coupling may give rise to variability in the frequencies
and amplitudes over time \citep{buchler1997}.  Given that this is statistically
the best model, and also physically the more plausible one, we focus on the
modes listed in Table\,\ref{tbl:combinations}, by performing a time-frequency
analysis for every prewhitening stage where the frequency under consideration is
the dominant one.

A logical approach is to perform a wavelet analysis, adapted to the
unequidistant signature of the dataset at hand \citep{foster1996}.  In a data
set with a low enough noise level, one can also compare the shape of the
detected peak $p(f)$ in the Fourier periodogram with the theoretical shape of an
infinitely stable mode of frequency $f_0$:
\[p(f) = \sqrt{\left(\frac{\sin[\pi T (f - f_0)]}{\pi T (f - f_0)}\right)^2},\]
with $T$ the total time span. This method has the advantage that we concentrate
on the most localized area possible in frequency space. On the downside, a peak
not fitting the expected shape can also mean that there is a beating pattern on
a longer time scale which is not well resolved by the data sets. A third method
to investigate stability we applied is simply to cut the entire time series in
half, and do an independent traditional Scargle analysis on both parts. 

\begin{figure*}
\centering\includegraphics[width=2\columnwidth]{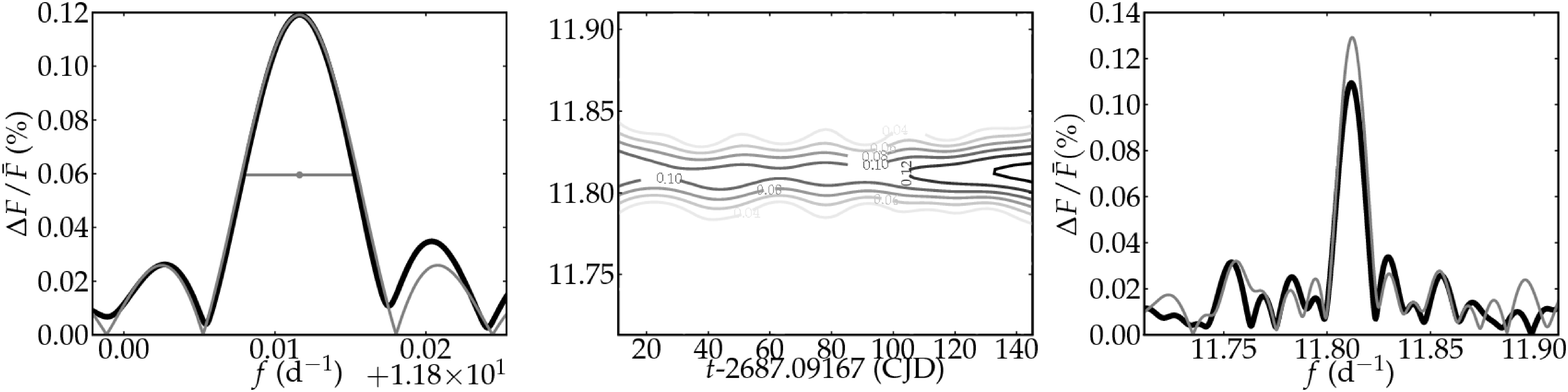}
\centering\includegraphics[width=2\columnwidth]{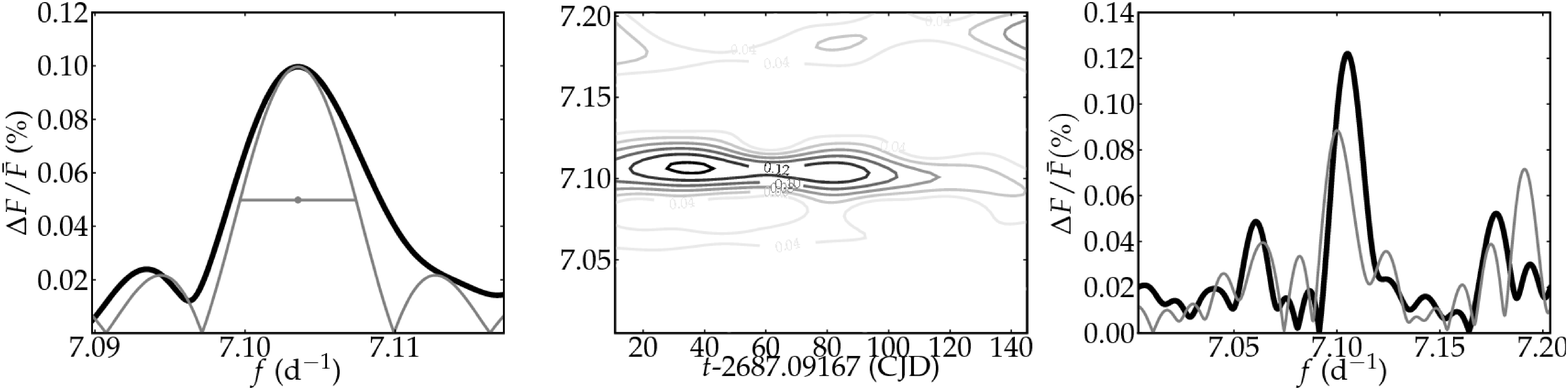}
\centering\includegraphics[width=2\columnwidth]{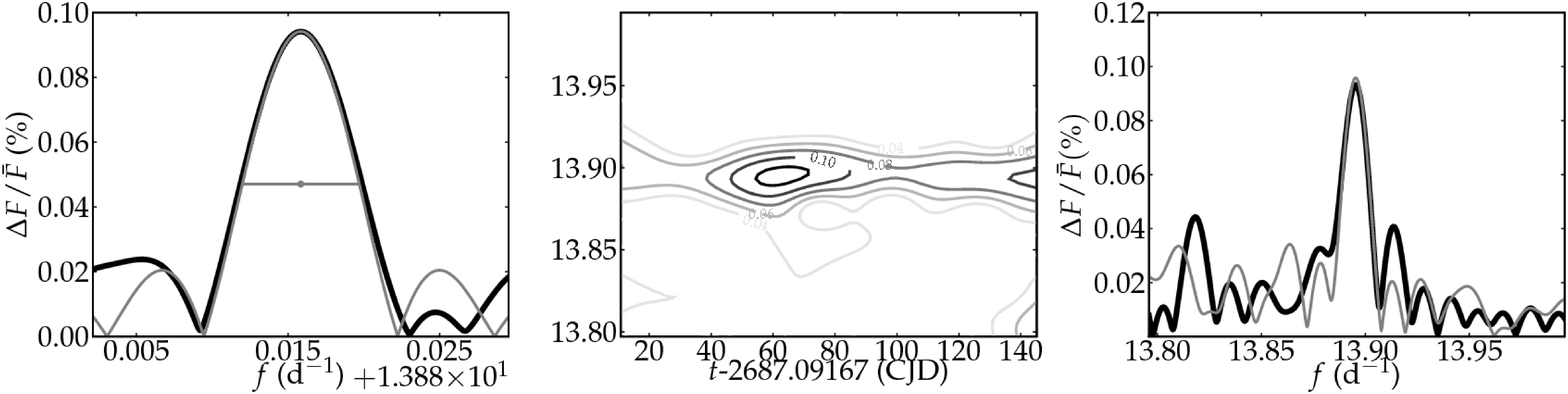}
\centering\includegraphics[width=2\columnwidth]{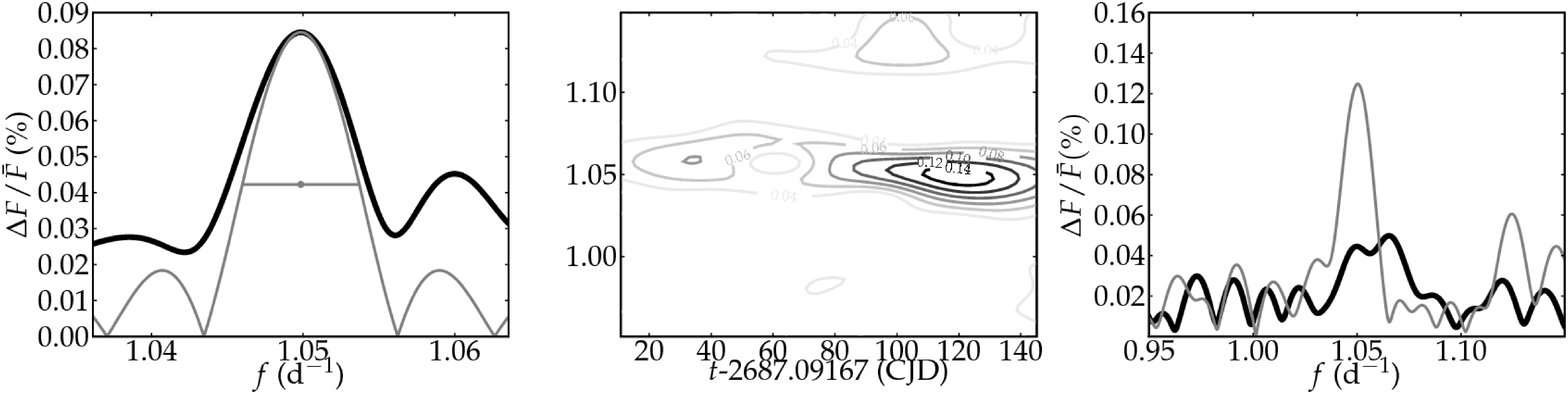}
\centering\includegraphics[width=2\columnwidth]{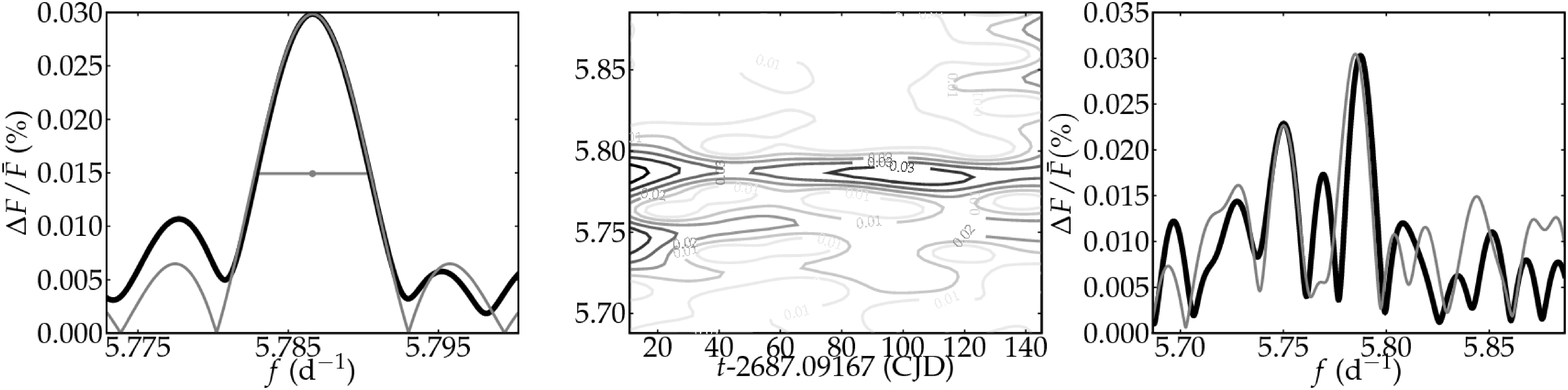}

\caption{Search for changes in the frequency or amplitude behaviour (in \% of
mean flux) for some of the frequencies in Table\,\ref{tbl:combinations}. Left
panels compare the observed shape of the Scargle peak (\emph{black}) with the
theoretical infinitely stable peak (\emph{gray}), middle panels show the result
of a wavelet analysis, right panels show the comparison between the peak as
calculated only using the first half of the time series (\emph{black}) and the
peak using only the last half (\emph{gray}). From top to bottom: $d_{1,3}(1,1)$
shows stability of the frequency, but a decrease in amplitude towards the end of
the time series; $f_7$ shows signs of a decrease in both amplitude and
frequency; $d_{1,4}(1,1)$ proves to be relatively stable; $d_{4,8}(1,-1)$ shows
a huge rise in amplitude from virtually non existent to highly significant;
$d_{1,2}(1,1)$ could evolve slowly in frequency.}\label{fig:varfreqs}
\end{figure*}

The results for some of the frequencies in Table\,\ref{tbl:combinations} are
shown in Fig.\,\ref{fig:varfreqs}, where the window for the wavelet transforms
was taken as 40 days around the targeted frequencies. The dominant mode
frequency and its first harmonic does not change during the entire CoRoT time
series.  Comparing the shapes of all the other peaks (some of which shown in the
left panels of Fig.\,\ref{fig:varfreqs}) leads to the conclusion that some
frequencies do not change their behaviour while others do.  This is particularly
the case for $d_{4,8}(1,-1)\approx1.05$\,d$^{-1}$.  The wavelet analysis hints
towards changes in the amplitudes of many of the modes, but it does not
allow thorough quantitative conclusions. Strong amplitude changes have also been found
in the CoRoT data of the pulsating Be star HD\,49330 \citep{huat2009}.

\section{Long term frequency evolution of the dominant mode}

\begin{figure}
\includegraphics[width=\columnwidth]{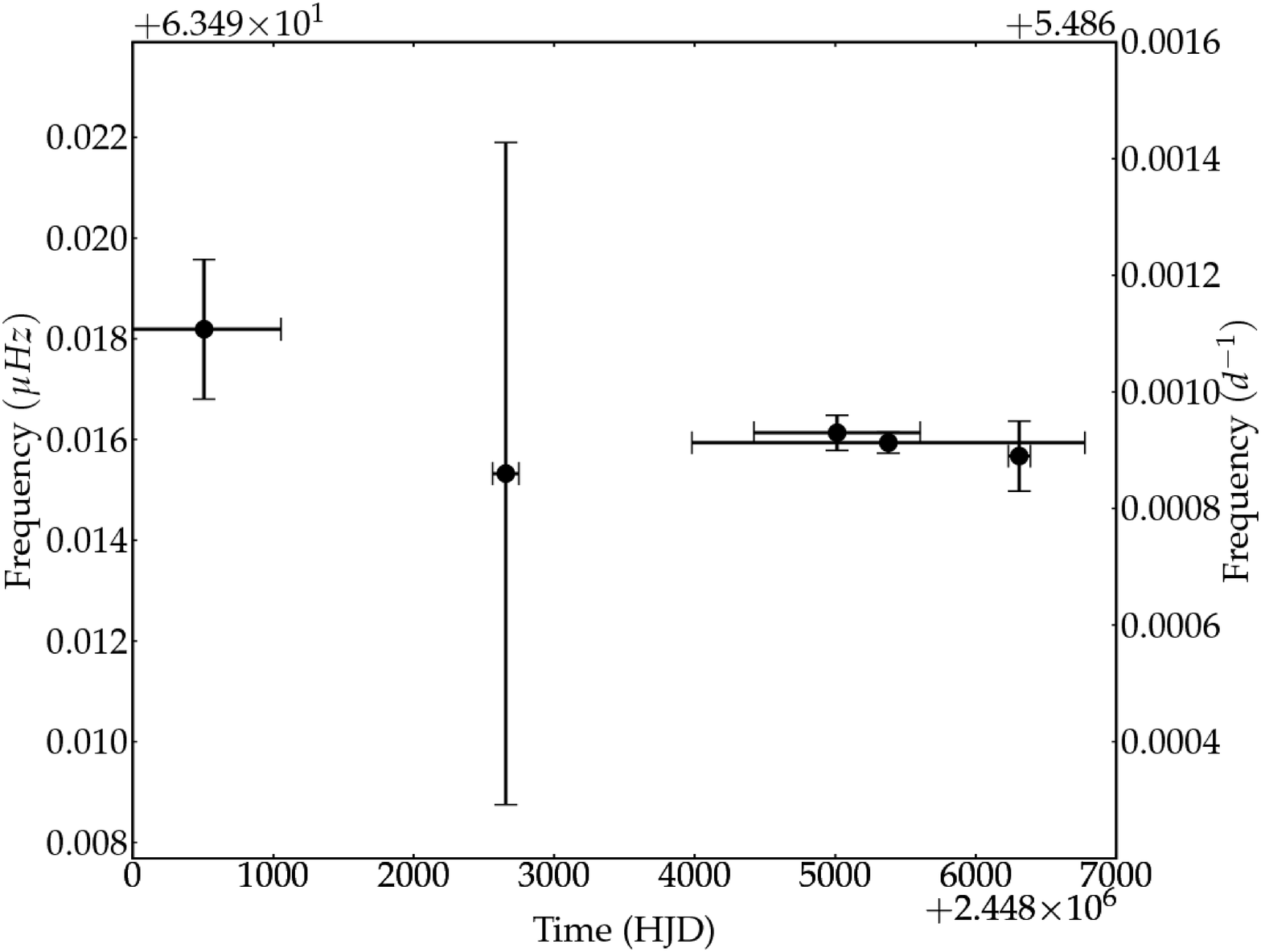}
\caption{Frequency determinations of ground and space based observations
(\emph{left to right:} Hipparcos, Swiss 70cm, Mercator, ASAS and
CoRoT). Vertical bars denote $2\sigma$ error in frequency, corrected for
correlation. Horizontal bars denote time span for the frequency determination,
circles denote midpoint of observations.}\label{fig:freqsummary}
\end{figure}

The Hipparcos satellite observed HD\,180642 from March 12, 1990 during almost
three years. Since then, 191 observations assembled with the photomultiplier P7
attached to the 0.7m Swiss telescope at La Silla and to the 1.2m Mercator
telescope at La Palma were added.  Moreover, we downloaded 310 archival ASAS
data points \citep{pigulski2008}.  This brings the
total time span of observations to 6814 days. The characteristics
of the different datasets, as well as the first frequency value determined for each
of these data sets are summarized in
Table\,\ref{tbl:ground_based} and Fig.\,\ref{fig:freqsummary}.

\begin{table*}
\caption{Datasets described in this paper, and frequency values for the dominant
mode determined from a Scargle periodogram (Scargle 1982).  All frequency values
were also calculated via a PDM procedure (Stellingwerf 1982), a multiharmonic
periodogram \citep{schwarzenberg1996} and a nonlinear least squares fit with 5
(fixed) harmonics, and were consistent with each other within 1\,$\sigma$. By
$t_0$ we indicate the starting time of the observations, while $T$ is the total
timespan, $n$ is the number of measurements, and $f$ and $\sigma_f$ are the
frequency and error, respectively).}  \centering\begin{tabular}{lcrrlll}
\hline\hline & $t_0$ (y) & $T$~(d) & $n$ & $f$ (d$^{-1}$) & $\sigma_f$
(d$^{-1}$) & $f$ ($\mu$Hz)\\\hline
Hipparcos     & 1990 & 1091 & 83       & 5.4871  & 0.00006  & 63.508(2)\\
Swiss 70cm/P7 & 1997 & 187  & 20       & 5.4869  & 0.0002  & 63.50(5)\\
Mercator/P7   & 2002 & 1184 & 171      & 5.48693 & 0.00001 & 63.506(1)\\
ASAS          & 2001 & 2798 & 310      & 5.48691 & 0.00001 & 63.505(9)\\
CoRoT/SISMO   & 2007 & 156  & 379\,785 & 5.48689 & 0.00003 & 63.505(7)\\ \hline
All           & 1990 & 6814 & 380\,369 & 5.48694 & 0.00003 & 63.506(3)\\ \hline\end{tabular}
\label{tbl:ground_based}
\end{table*}

The estimator (\ref{eq:frequency_error}) of the error on the frequency suggests
that the precision simply scales with the total time span $T$, which would hypothetically lower
the frequency error by several magnitudes in our case if the datasets were to be combined.
However, we have to take into account the extremely biased distribution of
observations in time: more than $99\%$ of the observations are made during
a time span only covering less than $3\%$ of the total time span. A natural measure
for the uncertainty on the frequency is given by the width of the peak in the window
function around $f=0$. When adding only 83 datapoints to nearly half a million measurements,
only tiny `wobbles' are added to the window function and the overall shape of the main
peaks is unaltered.

Instead of combining the datasets to model the stability of the frequency, we
treat the datasets separately, and test if the frequency derived from Hipparcos measurements are
 equal to the one derived from the CoRoT mission, within estimated errors.
In the CoRoT dataset correlation
effects are taken into account. To determine how accurate the error estimation is in the case of the Hipparcos data,
we randomly drew $\sim$7000 samples from the CoRoT dataset, using the (scaled) time gaps 
from the Hipparcos measurements, to arrive at a comparable number of datapoints
($\approx 100$) in each sample. We derive an empirical frequency error estimate of
$\hat{\sigma_f} = 0.0002$\,d$^{-1}$, while formula (\ref{eq:frequency_error}) gives us a conservatively overestimated
average value of $\hat{\sigma_f} = 0.001$\,d$^{-1}$.

If the frequencies for the dominant mode from Hipparcos and CoRoT are estimates for a common mean $\bar{f}$, then the maximum likelihood estimator for $\bar{f}$ is $\bar{f}=5.48691$\,d$^{-1}$, with a probability
of $0.6\%$. These facts suggest that the frequency of the dominant mode is not the same in these two data sets, but has instead decreased. From the datasets in Table\,\ref{tbl:ground_based},
it is impossible to determine if the frequency change is gradual or sudden, so we refrain from any physical interpretation, although it would
be naturally explained as a frequency decrease due to the approaching of the star to the end of the core-hydrogen burning, as suggested by the low $\log g$ of 3.45 dex.

\section{Residual power}

\begin{figure*}
\centering\includegraphics[width=\columnwidth]{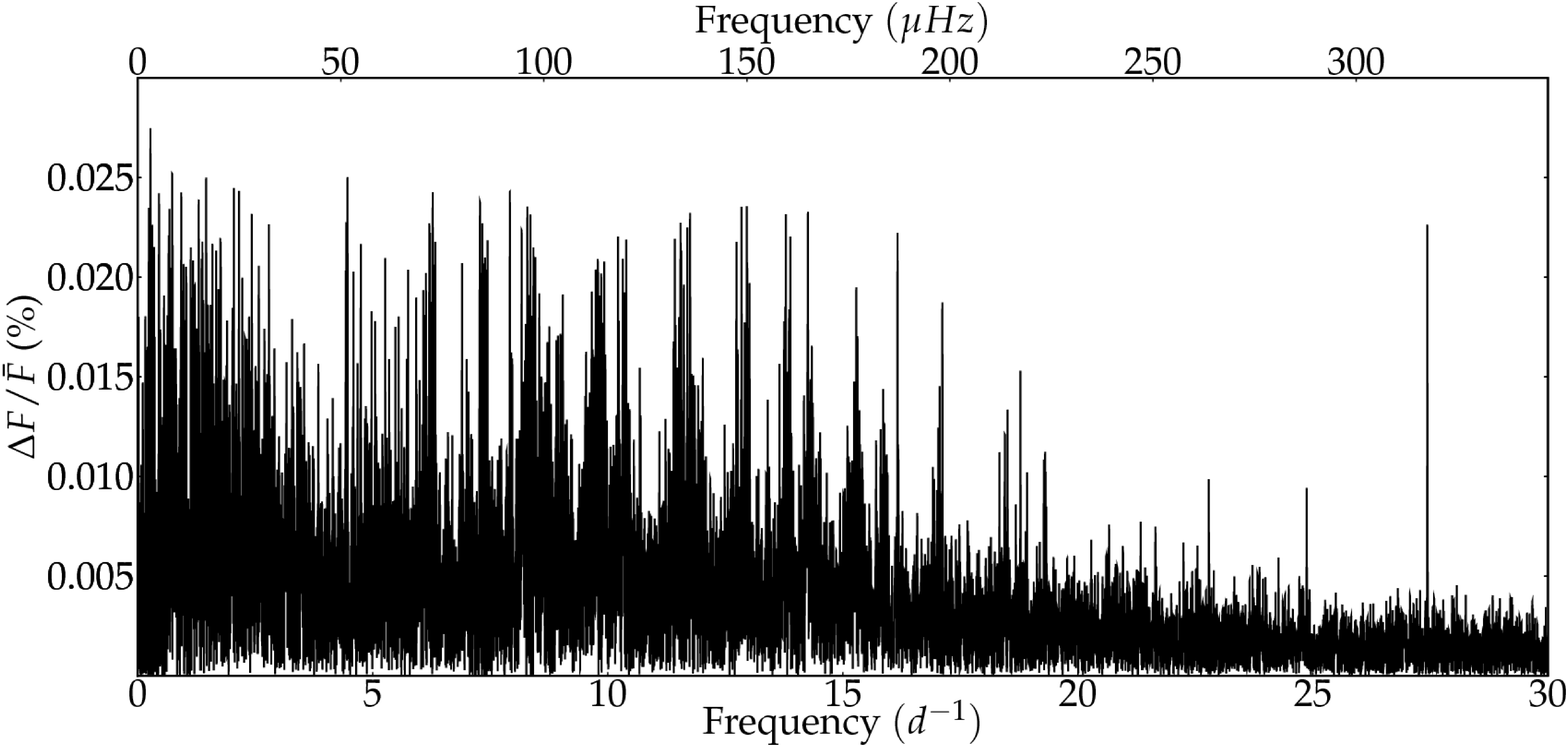}
\centering\includegraphics[width=\columnwidth]{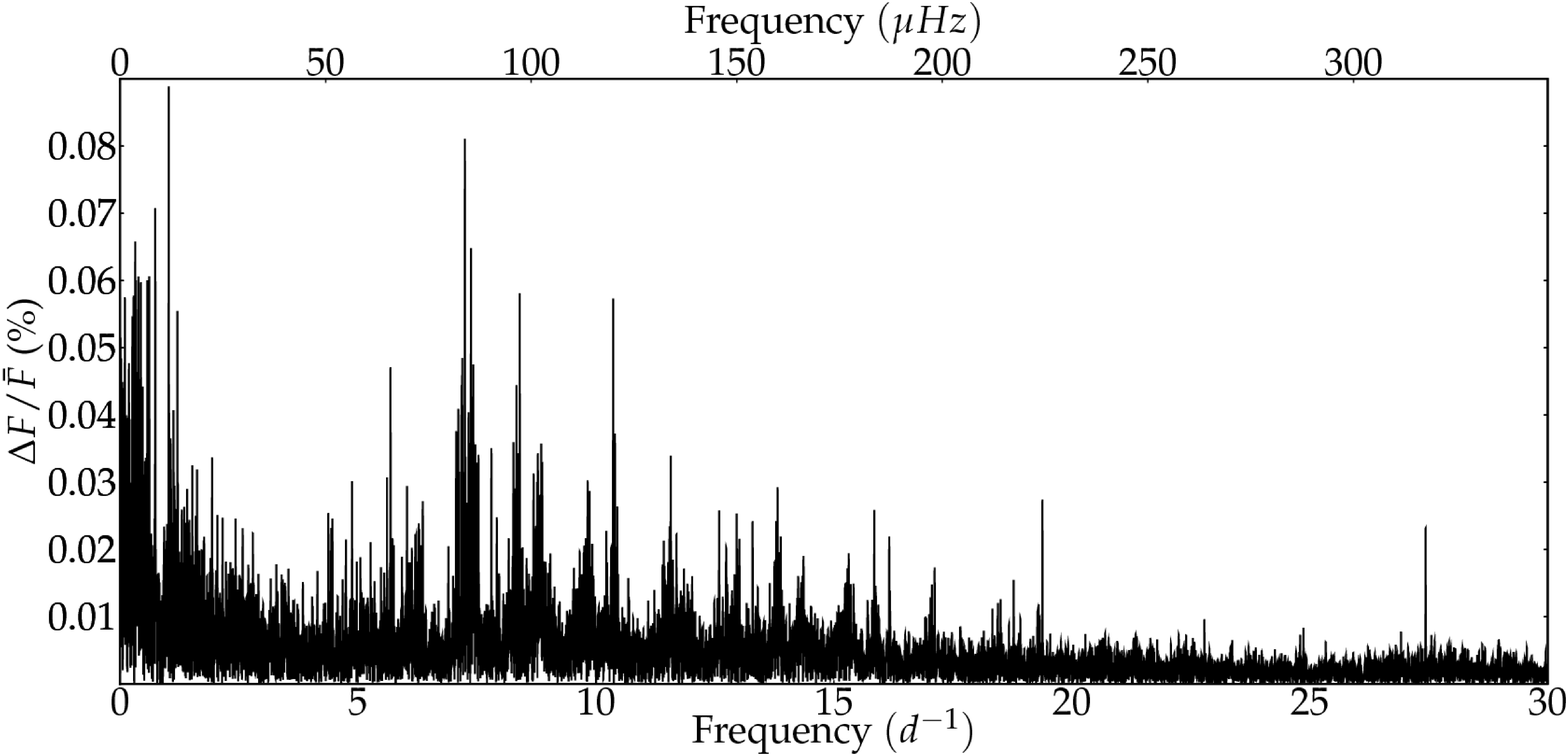}
\centering\includegraphics[width=\columnwidth]{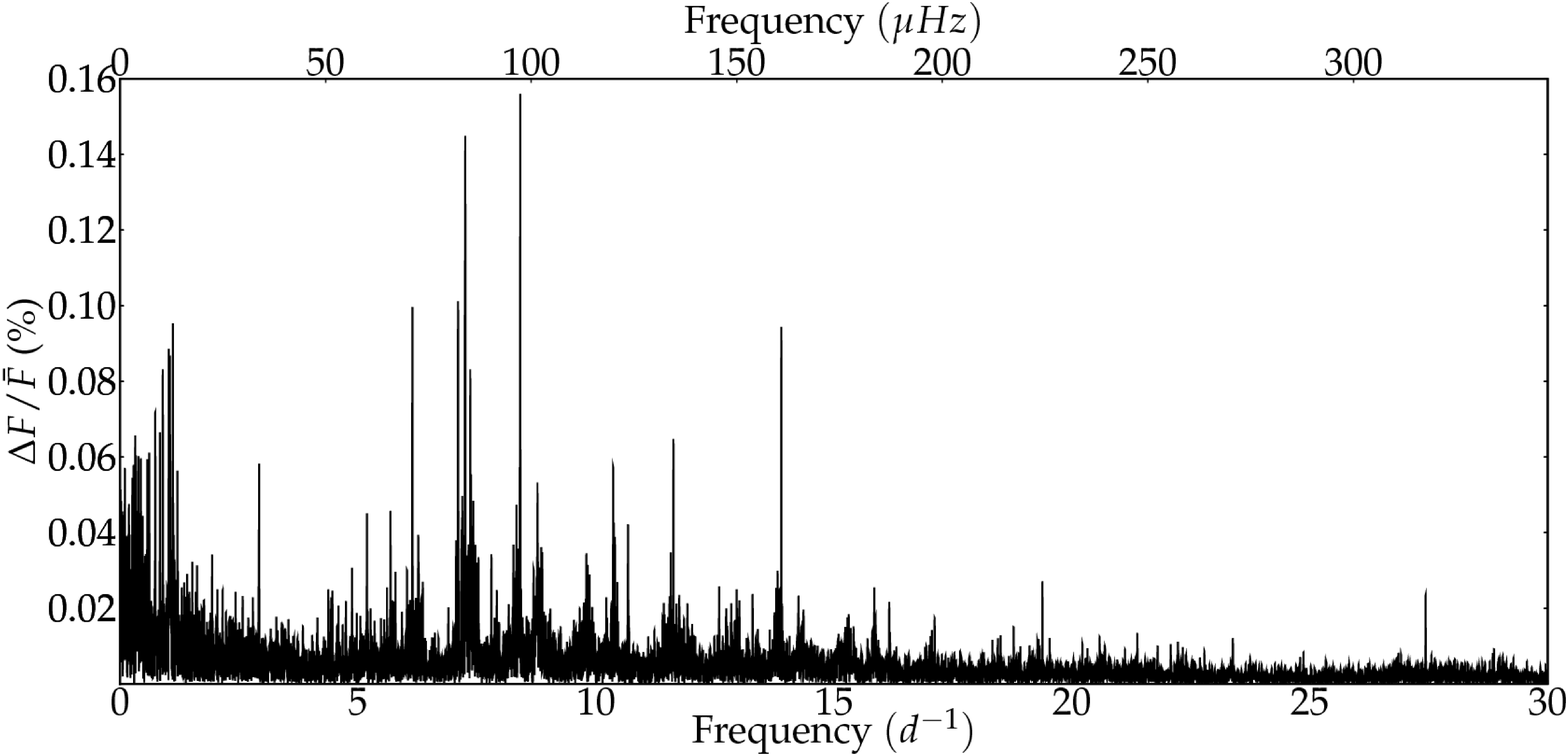}
\caption{Residual periodograms for each of the three prewhitened models $F_1,
F_2$ and $F_3$ described in the text.}
\label{residues}
\end{figure*}

As already hinted at above when discussing the three models $F_1, F_2$ and
$F_3$, we did not yet reach the noise level when considering all the 127
frequencies listed in Table\,\ref{raw_scargle}, or the more restricted lists
belonging to models $F_2$ and $F_3$ given in Tables\,\ref{tbl:var_9} to
\ref{tbl:var_10}. Figure\,\ref{residues} shows the residual periodograms for
each of the three considered models, where we took the time-variable amplitude
and phase model $F_3$ allowing for one frequency to describe this amplitude
variability as this leads to the best BIC (Table\,\ref{tbl:model_evaluation}),
but the result is similar for the other three cases of this model.
As can be seen, all three models lead to residual power excess, although at a
quite different level. Further prewhitening according to model $F_1$ was done
(see Table\,\ref{raw_scargle} which lists up to 200 frequencies) but is,
according to us, not very useful for a physical interpretation of the
frequencies. 

Recently, \citet{belkacem2009} interpreted the residual power of HD\,180642
after prewhitening 91 frequencies with a model description as $F_1$, in terms of
stochastically excited modes due to turbulent convection. They derived a large
spacing $\Delta\nu = 13.5\mu$Hz or twice this value, from the residual power
spectrum, excluding the frequencies below 130$\mu$Hz (11.23\,d$^{-1}$) and above
300$\mu$Hz (26\,d$^{-1}$).  We refer to their paper for a physical
description and interpretation of such modes.

Our model comparison shows that the residual power is quite different for our
preferred physical model $F_2$ than for model $F_1$ considering 127 frequencies.
Even though model $F_1$ leads to lower residual power, it is statistically less
good than model $F_2$ if one takes into account the difference in degrees of
freedom. Also, `natural' combination frequencies occur among the found
frequencies, of which several are phase-locked.  One may then wonder how
frequencies which are resonantly excited and which may show time-dependent
behaviour can be distinguished from stochastically excited ones, when they occur
in the same frequency regime.  In any case, the frequencies involved in phase
locking are not expected to have a stochastic nature, as they would have random
phase behaviour. Therefore, all the combination frequencies whose phases are
locked (6 sum and 6 difference frequencies --- see Fig.\,\ref{fig:combinations})
are likely not due to a stochastic process. They cover almost the entire range
in frequency covered in Fig.\,\ref{bars}.

We estimated the large separations for the three residual power spectra shown in
Fig.\,\ref{residues}, assuming them to be caused by stochastically excited
oscillations (p-modes), by using \'echelle diagrams with extracted frequencies
and autocorrelations \citep{christensen1988}.  We thus find, in the frequency range
from 50\,$\mu$Hz to 300\,$\mu$Hz, 
$\Delta\nu_1=12.1\pm0.2$\,$\mu$Hz for the residuals of model $F_1$,
$\Delta\nu_2=12.9\pm0.2$\,$\mu$Hz for the residuals of model $F_2$ and
$\Delta\nu_3=18.0\pm0.2$\,$\mu$Hz for the residuals of model $F_3$.  
The autocorrelation diagram for the
residuals of the models also shows a smaller bump around 24\,$\mu$Hz and
36\,$\mu$Hz.  Assuming $\ell=1$ modes as the cause of $\Delta\nu_1$, 
$\Delta\nu_3$ could be interpreted as the
distance between $\ell=0$ and $\ell=1$ modes but we regard this as a tentative
result which needs further confirmation.

The detection of solar-like oscillations in this $\beta\,$Cep star as an
explanation of the residual excess power seems to be robust against the various
models for the prewhitening of the $\kappa$-driven and resonantly excited modes.
The value of the large spacing is, however, light curve model dependent.

\section{Discussion and conclusion}

The available CoRoT data of the $\beta\,$Cep star HD\,180642 provided us with a
wealth of information.

Beyond the previously known dominant mode, many more frequencies are detected. A
large fraction of those does not change their behaviour during the time span of
the CoRoT data, while others do.  Light curve modelling using different
underlying functional assumptions led us to prefer a model based on nonlinear
mode interaction, with 11 independent frequencies and 22 three-resonance
combinations (among which some harmonics) covering the frequency range
$[0.3;22]$\,d$^{-1}$.  This model selection was based on statistical criteria,
without considering physical arguments {\it a priori}.  Nine of the independent
frequencies of this model are in the range expected for $\beta\,$Cep stars,
i.e., between 5 and 9\,d$^{-1}$.  This model is also the most logical one in
terms of physical interpretation. Indeed, the nonlinear frequency locking is a
natural consequence of the large amplitude of the dominant radial mode of the
star. Five of these 33 frequencies are in the regime of high-order g modes with
frequencies below 2\,d$^{-1}$ for stellar models appropriate for the star. The
relative amplitudes of the coupling frequencies differ an order of magnitude and
seem to point towards nonlinear resonant mode excitation and phase locking for
several of these frequencies, particularly for those in the g-mode frequency
regime.

Our observational results constitute a fruitful starting point for
detailed seismic modelling of this star, particularly if some of the frequencies
derived here could be identified. An extensive ground-based observing campaign
has been organised with that goal and is discussed in \citet{briquet2009}. As
indicated on Fig.\,\ref{bars}, it leads to fully consistent frequency results
with those found here, with 9 high-amplitude frequencies in common.

\begin{acknowledgements}
The research leading to these results has received funding from the European
Research Council under the European Community's Seventh Framework Programme
(FP7/2007--2013)/ERC grant agreement n$^\circ$227224 (PROSPERITY), as well as
from the Research Council of K.U.Leuven grant agreement GOA/2008/04 and from the
Belgian PRODEX Office under contract C90309: CoRoT Data Exploitation. This work
was supported by the italian ESS project, contract ASI/INAF I/015/07/0,
WP\,03170. KU acknowledges financial support from a \emph{European Community
Marie Curie Intra-European Fellowship}, contract number MEIF-CT-2006-024476.
\end{acknowledgements}

\bibliographystyle{aa}
\bibliography{aa_HD180642_version8}

\Online
\begin{appendix}

\section{Frequency tables}
\footnotesize
\addtocounter{table}{1} 
\longtab{1}{\begin{longtable}{rrrrrrrrrrrrr}
\caption{\label{raw_scargle}Phase $\phi$ in $2\pi/{\rm rad}$, variance reduction
  (VR), AIC and BIC in percentage. Amplitudes in fraction of the mean observed flux. Signal-to-noise (SNR) is calculated over a
  6\,d$^{-1}$ in the periodogram; $C=-0.00005$.}
\\
\hline\hline
$n$&$f$ (d$^{-1}$)&$\epsilon_f$ (d$^{-1}$)&$f$ ($\mu$Hz)&$\epsilon_f$ ($\mu$Hz)&$\Delta A/\bar{F}$&$\epsilon_A$&$\phi$&$\epsilon_\phi$&SNR&VR&AIC&BIC\\

\hline
\endfirsthead
\caption{continued.}\\
\hline\hline
$n$&$f$ (d$^{-1}$)&$\epsilon_f$ (d$^{-1}$)&$f$ ($\mu$Hz)&$\epsilon_f$ ($\mu$Hz)&$\Delta A/\bar{F}$&$\epsilon_A$&$\phi$&$\epsilon_\phi$&SNR&VR&AIC&BIC\\
\hline
\endhead
\hline
\endfoot
1&5.486889&0.000023&63.505662&0.000262&0.03696&0.00024&-0.0355&0.0064&188.7&91.86&100.00&100.00\\ 
2&10.973740&0.000062&127.010883&0.000721&0.00631&0.00011&-0.4938&0.0177&63.0&94.54&98.24&98.18\\ 
3&16.460665&0.000080&190.516954&0.000928&0.00405&0.00009&0.3492&0.0227&70.4&95.65&97.25&97.15\\ 
4&0.299171&0.000093&3.462623&0.001073&0.00313&0.00008&-0.3748&0.0263&27.8&96.31&96.53&96.41\\ 
5&6.324816&0.000120&73.203886&0.001389&0.00224&0.00008&-0.2335&0.0340&21.8&96.64&96.11&95.98\\ 
6&8.409185&0.000161&97.328527&0.001864&0.00156&0.00007&0.4594&0.0457&17.9&96.81&95.89&95.77\\ 
7&7.254757&0.000162&83.967094&0.001875&0.00145&0.00007&-0.0600&0.0460&16.5&96.95&95.70&95.58\\ 
8&21.947560&0.000184&254.022686&0.002130&0.00127&0.00007&0.1164&0.0522&36.6&97.06&95.54&95.42\\ 
9&11.811640&0.000193&136.708796&0.002232&0.00119&0.00007&-0.4081&0.0547&15.4&97.16&95.40&95.29\\ 
10&6.143358&0.000233&71.103681&0.002696&0.00100&0.00007&-0.0105&0.0661&10.1&97.22&95.30&95.19\\ 
11&7.103534&0.000224&82.216823&0.002595&0.00100&0.00006&0.3475&0.0636&11.5&97.29&95.19&95.10\\ 
12&1.112160&0.000234&12.872226&0.002710&0.00095&0.00006&-0.1078&0.0664&9.4&97.35&95.09&95.01\\ 
13&13.895848&0.000235&160.831579&0.002716&0.00094&0.00006&0.2959&0.0666&14.9&97.41&95.00&94.92\\ 
14&1.024105&0.000240&11.853066&0.002775&0.00089&0.00006&-0.4860&0.0680&8.9&97.46&94.91&94.84\\ 
15&7.358664&0.000248&85.169719&0.002874&0.00085&0.00006&0.3891&0.0705&10.2&97.51&94.83&94.77\\ 
16&1.049850&0.000240&12.151046&0.002779&0.00084&0.00006&0.2368&0.0681&8.6&97.56&94.75&94.69\\ 
17&0.898703&0.000246&10.401656&0.002843&0.00084&0.00006&0.1240&0.0697&8.5&97.61&94.66&94.62\\ 
18&7.245672&0.000233&83.861941&0.002697&0.00085&0.00006&0.1250&0.0661&9.7&97.65&94.58&94.55\\ 
19&0.741254&0.000280&8.579333&0.003243&0.00071&0.00006&-0.4799&0.0795&7.2&97.69&94.52&94.50\\ 
20&0.320416&0.000287&3.708515&0.003325&0.00067&0.00005&-0.4893&0.0815&6.6&97.72&94.47&94.45\\ 
21&7.374444&0.000287&85.352356&0.003321&0.00066&0.00005&-0.3128&0.0814&7.9&97.74&94.42&94.41\\ 
22&11.630388&0.000287&134.610967&0.003326&0.00065&0.00005&-0.1838&0.0815&8.9&97.77&94.37&94.37\\ 
23&0.837937&0.000290&9.698339&0.003362&0.00064&0.00005&-0.4509&0.0824&6.7&97.80&94.31&94.33\\ 
24&0.573389&0.000304&6.636450&0.003514&0.00061&0.00005&0.0670&0.0861&6.3&97.83&94.27&94.29\\ 
25&0.613211&0.000297&7.097346&0.003440&0.00062&0.00005&0.3570&0.0843&6.5&97.85&94.22&94.25\\ 
26&0.437363&0.000304&5.062076&0.003515&0.00060&0.00005&-0.2705&0.0862&6.2&97.88&94.17&94.22\\ 
27&0.387200&0.000306&4.481479&0.003547&0.00059&0.00005&0.3086&0.0869&6.1&97.90&94.13&94.18\\ 
28&2.921596&0.000304&33.814774&0.003514&0.00058&0.00005&-0.4474&0.0861&6.0&97.92&94.08&94.15\\ 
29&10.361777&0.000302&119.927974&0.003496&0.00058&0.00005&-0.3554&0.0857&7.7&97.94&94.04&94.12\\ 
30&0.100806&0.000308&1.166736&0.003565&0.00058&0.00005&-0.0958&0.0874&6.1&97.97&93.99&94.08\\ 
31&8.396265&0.000307&97.178990&0.003548&0.00057&0.00005&0.2305&0.0870&7.3&97.99&93.95&94.05\\ 
32&1.205832&0.000308&13.956386&0.003563&0.00056&0.00005&-0.4130&0.0873&6.1&98.01&93.91&94.02\\ 
33&0.285145&0.000320&3.300292&0.003706&0.00055&0.00005&-0.2305&0.0908&5.8&98.03&93.87&93.99\\ 
34&8.770865&0.000326&101.514639&0.003768&0.00053&0.00005&-0.0214&0.0924&6.9&98.05&93.83&93.96\\ 
35&0.335229&0.000323&3.879971&0.003743&0.00053&0.00005&0.4316&0.0917&5.6&98.07&93.79&93.93\\ 
36&0.257337&0.000327&2.978433&0.003783&0.00052&0.00005&-0.2497&0.0927&5.6&98.08&93.75&93.91\\ 
37&0.009469&0.000310&0.109600&0.003582&0.00055&0.00005&-0.0817&0.0878&5.6&98.10&93.72&93.88\\ 
38&0.370980&0.000332&4.293751&0.003840&0.00050&0.00005&-0.0996&0.0941&5.3&98.12&93.68&93.86\\ 
39&7.425424&0.000350&85.942402&0.004053&0.00047&0.00005&0.0174&0.0994&5.9&98.13&93.65&93.84\\ 
40&5.684955&0.000351&65.798086&0.004058&0.00047&0.00005&-0.1344&0.0995&5.3&98.15&93.62&93.82\\ 
41&0.187774&0.000349&2.173308&0.004043&0.00047&0.00005&-0.2518&0.0991&5.2&98.16&93.59&93.80\\ 
42&7.191284&0.000355&83.232453&0.004109&0.00046&0.00005&0.3525&0.1007&5.8&98.18&93.56&93.78\\ 
43&8.331328&0.000356&96.427411&0.004123&0.00046&0.00005&-0.4133&0.1011&6.0&98.19&93.53&93.76\\ 
44&0.091000&0.000354&1.053236&0.004102&0.00046&0.00005&-0.1169&0.1005&4.9&98.20&93.50&93.74\\ 
45&0.050942&0.000364&0.589602&0.004213&0.00044&0.00005&0.4170&0.1033&4.9&98.22&93.47&93.72\\ 
46&0.026107&0.000356&0.302160&0.004124&0.00045&0.00005&-0.4822&0.1011&4.9&98.23&93.44&93.71\\ 
47&5.187812&0.000373&60.044119&0.004317&0.00043&0.00005&0.4664&0.1058&4.9&98.24&93.42&93.69\\ 
48&10.674584&0.000369&123.548429&0.004270&0.00043&0.00004&-0.2976&0.1047&5.9&98.25&93.39&93.67\\ 
49&10.374926&0.000379&120.080167&0.004382&0.00042&0.00004&-0.2046&0.1074&5.7&98.26&93.36&93.66\\ 
50&7.109275&0.000377&82.283274&0.004361&0.00042&0.00004&0.0798&0.1069&5.2&98.28&93.34&93.65\\ 
51&8.368705&0.000372&96.860009&0.004309&0.00042&0.00004&-0.0134&0.1056&5.5&98.29&93.31&93.63\\ 
52&0.476009&0.000373&5.509363&0.004319&0.00042&0.00004&-0.3027&0.1059&4.8&98.30&93.28&93.62\\ 
53&1.122346&0.000370&12.990121&0.004288&0.00041&0.00004&0.1337&0.1051&4.7&98.31&93.26&93.60\\ 
54&6.265172&0.000393&72.513563&0.004549&0.00039&0.00004&0.0353&0.1115&4.7&98.32&93.23&93.59\\ 
55&17.298406&0.000393&200.213032&0.004550&0.00039&0.00004&0.3315&0.1115&9.1&98.33&93.21&93.58\\ 
56&0.130294&0.000397&1.508027&0.004589&0.00039&0.00004&-0.1917&0.1125&4.5&98.34&93.19&93.57\\ 
57&7.323153&0.000399&84.758713&0.004618&0.00038&0.00004&-0.0064&0.1132&5.0&98.35&93.16&93.56\\ 
58&7.064826&0.000402&81.768818&0.004648&0.00038&0.00004&-0.4697&0.1139&4.8&98.36&93.14&93.54\\ 
59&7.177056&0.000398&83.067776&0.004602&0.00038&0.00004&0.4879&0.1128&4.9&98.37&93.12&93.53\\ 
60&0.419303&0.000398&4.853043&0.004602&0.00038&0.00004&0.4717&0.1128&4.4&98.38&93.10&93.52\\ 
61&8.266944&0.000408&95.682225&0.004721&0.00037&0.00004&-0.1626&0.1157&5.0&98.39&93.08&93.51\\ 
62&0.068790&0.000393&0.796186&0.004548&0.00038&0.00004&0.2876&0.1115&4.3&98.40&93.05&93.50\\ 
63&0.598913&0.000400&6.931862&0.004624&0.00037&0.00004&0.2735&0.1133&4.4&98.41&93.03&93.49\\ 
64&4.648449&0.000410&53.801491&0.004747&0.00036&0.00004&0.2154&0.1163&4.2&98.42&93.01&93.48\\ 
65&1.062249&0.000405&12.294552&0.004689&0.00036&0.00004&-0.1036&0.1149&4.4&98.42&92.99&93.47\\ 
66&7.530514&0.000407&87.158725&0.004716&0.00036&0.00004&0.1886&0.1156&4.8&98.43&92.97&93.46\\ 
67&8.851947&0.000408&102.453088&0.004720&0.00036&0.00004&0.4543&0.1157&5.0&98.44&92.95&93.46\\ 
68&0.200097&0.000404&2.315934&0.004679&0.00036&0.00004&0.4588&0.1147&4.3&98.45&92.93&93.45\\ 
69&7.804933&0.000410&90.334876&0.004746&0.00035&0.00004&0.3967&0.1163&4.8&98.46&92.91&93.44\\ 
70&7.451400&0.000405&86.243060&0.004688&0.00035&0.00004&0.1652&0.1149&4.7&98.47&92.89&93.43\\ 
71&10.399457&0.000410&120.364087&0.004749&0.00035&0.00004&0.1639&0.1164&5.0&98.48&92.87&93.42\\ 
72&1.045241&0.000364&12.097692&0.004209&0.00039&0.00004&-0.1921&0.1032&4.2&98.48&92.85&93.41\\ 
73&11.573429&0.000417&133.951724&0.004829&0.00034&0.00004&0.2994&0.1184&5.1&98.49&92.83&93.41\\ 
74&8.779805&0.000400&101.618112&0.004630&0.00035&0.00004&-0.4466&0.1135&4.8&98.50&92.81&93.40\\ 
75&7.498775&0.000416&86.791379&0.004815&0.00034&0.00004&0.4624&0.1180&4.6&98.51&92.79&93.39\\ 
76&9.799902&0.000419&113.424788&0.004851&0.00034&0.00004&0.0974&0.1189&4.9&98.51&92.77&93.39\\ 
77&8.763891&0.000409&101.433921&0.004738&0.00034&0.00004&0.2360&0.1161&4.8&98.52&92.75&93.38\\ 
78&1.618521&0.000412&18.732877&0.004765&0.00033&0.00004&0.1506&0.1168&4.1&98.53&92.74&93.37\\ 
79&8.872841&0.000417&102.694924&0.004824&0.00033&0.00004&0.0555&0.1182&4.7&98.54&92.72&93.36\\ 
80&0.176488&0.000411&2.042685&0.004754&0.00033&0.00004&-0.0624&0.1165&4.0&98.54&92.70&93.36\\ 
81&1.935029&0.000424&22.396164&0.004910&0.00032&0.00004&0.0156&0.1203&4.0&98.55&92.68&93.35\\ 
82&0.380644&0.000404&4.405601&0.004673&0.00033&0.00004&-0.3600&0.1145&3.9&98.56&92.66&93.35\\ 
83&7.417077&0.000408&85.845797&0.004719&0.00033&0.00004&0.3142&0.1157&4.4&98.56&92.65&93.34\\ 
84&1.520179&0.000427&17.594661&0.004939&0.00032&0.00004&-0.4267&0.1211&3.9&98.57&92.63&93.33\\ 
85&0.222217&0.000422&2.571956&0.004887&0.00032&0.00004&0.4562&0.1198&3.9&98.58&92.61&93.33\\ 
86&8.359820&0.000410&96.757173&0.004747&0.00033&0.00004&0.4324&0.1164&4.5&98.58&92.60&93.32\\ 
87&7.259464&0.000385&84.021572&0.004457&0.00035&0.00004&-0.1884&0.1092&4.3&98.59&92.58&93.32\\ 
88&7.380371&0.000419&85.420960&0.004849&0.00032&0.00004&-0.0678&0.1188&4.3&98.60&92.56&93.31\\ 
89&7.479556&0.000421&86.568930&0.004873&0.00032&0.00004&0.2767&0.1194&4.4&98.60&92.54&93.31\\ 
90&4.874065&0.000442&56.412785&0.005113&0.00030&0.00004&-0.3730&0.1253&3.8&98.61&92.53&93.30\\ 
91&8.688025&0.000442&100.555841&0.005119&0.00030&0.00004&-0.3248&0.1255&4.4&98.62&92.51&93.30\\ 
92&5.609340&0.000434&64.922921&0.005029&0.00030&0.00004&0.2954&0.1233&3.9&98.62&92.50&93.30\\ 
93&9.825918&0.000436&113.725898&0.005042&0.00030&0.00004&-0.1101&0.1236&4.5&98.63&92.48&93.29\\ 
94&0.430350&0.000426&4.980899&0.004934&0.00031&0.00004&-0.1086&0.1209&3.7&98.63&92.46&93.29\\ 
95&1.070228&0.000420&12.386895&0.004862&0.00031&0.00004&-0.4757&0.1192&3.9&98.64&92.45&93.28\\ 
96&5.786624&0.000435&66.974815&0.005032&0.00030&0.00004&0.2157&0.1233&3.9&98.65&92.43&93.28\\ 
97&6.031242&0.000438&69.806041&0.005066&0.00030&0.00004&0.4529&0.1242&3.9&98.65&92.42&93.28\\ 
98&7.123896&0.000428&82.452503&0.004953&0.00030&0.00004&-0.3384&0.1214&4.1&98.66&92.40&93.27\\ 
99&13.818031&0.000443&159.930915&0.005129&0.00029&0.00004&0.3384&0.1257&5.2&98.66&92.39&93.27\\ 
100&0.038176&0.000434&0.441848&0.005022&0.00030&0.00004&0.3760&0.1231&3.8&98.67&92.37&93.27\\ 
101&0.409477&0.000407&4.739314&0.004708&0.00031&0.00004&-0.2513&0.1154&3.7&98.67&92.36&93.26\\ 
102&1.411052&0.000436&16.331623&0.005050&0.00029&0.00004&0.2190&0.1238&3.8&98.68&92.34&93.26\\ 
103&0.521680&0.000439&6.037968&0.005078&0.00029&0.00004&-0.3007&0.1245&3.8&98.69&92.33&93.25\\ 
104&7.221801&0.000436&83.585665&0.005043&0.00029&0.00004&0.2655&0.1236&4.1&98.69&92.31&93.25\\ 
105&0.114405&0.000440&1.324132&0.005092&0.00029&0.00004&0.0372&0.1248&3.7&98.70&92.30&93.25\\ 
106&0.228917&0.000444&2.649499&0.005136&0.00029&0.00004&-0.2926&0.1259&3.7&98.70&92.28&93.25\\ 
107&7.466602&0.000443&86.419001&0.005127&0.00029&0.00004&-0.1099&0.1257&4.0&98.71&92.27&93.24\\ 
108&13.291190&0.000452&153.833214&0.005236&0.00028&0.00004&0.3146&0.1283&4.9&98.71&92.25&93.24\\ 
109&8.807314&0.000449&101.936500&0.005200&0.00028&0.00004&-0.4451&0.1274&4.2&98.72&92.24&93.24\\ 
110&8.823592&0.000412&102.124902&0.004772&0.00031&0.00004&-0.3130&0.1170&4.5&98.72&92.22&93.23\\ 
111&8.837597&0.000434&102.287004&0.005019&0.00029&0.00004&-0.0539&0.1230&4.3&98.73&92.21&93.23\\ 
112&6.362414&0.000444&73.639052&0.005137&0.00028&0.00003&-0.1784&0.1259&3.8&98.73&92.19&93.23\\ 
113&8.302855&0.000444&96.097863&0.005135&0.00028&0.00003&-0.4654&0.1259&4.1&98.74&92.18&93.23\\ 
114&9.868206&0.000447&114.215342&0.005174&0.00027&0.00003&-0.4197&0.1268&4.2&98.74&92.16&93.22\\ 
115&19.382612&0.000449&224.335786&0.005202&0.00027&0.00003&0.0761&0.1275&7.5&98.75&92.15&93.22\\ 
116&8.735203&0.000451&101.101887&0.005221&0.00027&0.00003&0.0035&0.1280&4.1&98.75&92.14&93.22\\ 
117&0.137406&0.000446&1.590343&0.005163&0.00027&0.00003&0.2740&0.1266&3.5&98.76&92.12&93.22\\ 
118&8.376218&0.000444&96.946966&0.005136&0.00027&0.00003&0.1986&0.1259&4.0&98.76&92.11&93.22\\ 
119&1.143357&0.000460&13.233304&0.005327&0.00027&0.00003&-0.0138&0.1306&3.6&98.77&92.10&93.22\\ 
120&10.449163&0.000467&120.939385&0.005406&0.00026&0.00003&0.4923&0.1325&4.1&98.77&92.08&93.22\\ 
121&7.184473&0.000455&83.153622&0.005271&0.00027&0.00003&0.1653&0.1292&3.8&98.78&92.07&93.22\\ 
122&12.588851&0.000468&145.704299&0.005418&0.00026&0.00003&0.0927&0.1328&4.5&98.78&92.06&93.21\\ 
123&1.347183&0.000473&15.592394&0.005479&0.00026&0.00003&0.3051&0.1343&3.5&98.79&92.05&93.21\\ 
124&4.375634&0.000474&50.643915&0.005483&0.00026&0.00003&-0.2058&0.1344&3.5&98.79&92.03&93.21\\ 
125&15.849010&0.000472&183.437619&0.005461&0.00026&0.00003&0.4390&0.1339&5.7&98.79&92.02&93.21\\ 
126&0.542048&0.000473&6.273707&0.005470&0.00025&0.00003&-0.3701&0.1341&3.4&98.80&92.01&93.21\\ 
127&0.989224&0.000468&11.449349&0.005411&0.00025&0.00003&0.4361&0.1326&3.5&98.80&92.00&93.21\\ 
128&4.463677&0.000470&51.662926&0.005437&0.00025&0.00003&0.1705&0.1333&3.4&98.81&91.98&93.21\\ 
129&1.457477&0.000474&16.868952&0.005483&0.00025&0.00003&-0.2285&0.1344&3.4&98.81&91.97&93.21\\ 
130&0.270580&0.000468&3.131714&0.005412&0.00025&0.00003&-0.1148&0.1327&3.3&98.82&91.96&93.21\\ 
131&7.917871&0.000482&91.642025&0.005580&0.00025&0.00003&-0.2305&0.1368&3.7&98.82&91.95&93.21\\ 
132&1.296026&0.000483&15.000305&0.005589&0.00025&0.00003&-0.4679&0.1370&3.4&98.82&91.94&93.22\\ 
133&0.237799&0.000454&2.752299&0.005256&0.00026&0.00003&-0.3142&0.1288&3.3&98.83&91.93&93.22\\ 
134&2.043672&0.000478&23.653608&0.005535&0.00025&0.00003&-0.0814&0.1357&3.5&98.83&91.92&93.22\\ 
135&0.736254&0.000447&8.521458&0.005169&0.00026&0.00003&0.1303&0.1267&3.4&98.84&91.90&93.22\\ 
136&0.924792&0.000474&10.703616&0.005483&0.00024&0.00003&0.3061&0.1344&3.4&98.84&91.89&93.22\\ 
137&6.274253&0.000455&72.618664&0.005271&0.00025&0.00003&-0.0798&0.1292&3.5&98.84&91.88&93.22\\ 
138&0.453644&0.000455&5.250507&0.005268&0.00025&0.00003&0.1580&0.1291&3.4&98.85&91.87&93.22\\ 
139&8.352350&0.000463&96.670712&0.005356&0.00025&0.00003&0.2023&0.1313&3.7&98.85&91.86&93.22\\ 
140&1.128681&0.000462&13.063439&0.005343&0.00025&0.00003&0.4418&0.1310&3.4&98.86&91.85&93.22\\ 
141&2.157032&0.000474&24.965648&0.005490&0.00024&0.00003&0.4815&0.1346&3.5&98.86&91.83&93.22\\ 
142&14.257403&0.000477&165.016241&0.005526&0.00024&0.00003&-0.1181&0.1355&4.6&98.86&91.82&93.22\\ 
143&1.038144&0.000470&12.015553&0.005439&0.00024&0.00003&-0.2466&0.1333&3.4&98.87&91.81&93.22\\ 
144&2.427459&0.000484&28.095586&0.005598&0.00023&0.00003&-0.1941&0.1372&3.4&98.87&91.80&93.22\\ 
145&0.347300&0.000475&4.019673&0.005495&0.00024&0.00003&-0.2060&0.1347&3.3&98.87&91.79&93.22\\ 
146&11.751733&0.000487&136.015424&0.005638&0.00023&0.00003&-0.1587&0.1382&4.0&98.88&91.78&93.23\\ 
147&7.278539&0.000482&84.242348&0.005576&0.00024&0.00003&-0.4180&0.1367&3.5&98.88&91.77&93.23\\ 
148&7.333070&0.000459&84.873501&0.005313&0.00025&0.00003&-0.1156&0.1302&3.5&98.88&91.76&93.23\\ 
149&12.955309&0.000487&149.945711&0.005633&0.00023&0.00003&0.4954&0.1381&4.2&98.89&91.75&93.23\\ 
150&12.844315&0.000487&148.661057&0.005641&0.00023&0.00003&0.1819&0.1383&4.2&98.89&91.74&93.23\\ 
151&6.201708&0.000483&71.779026&0.005588&0.00023&0.00003&-0.2326&0.1370&3.4&98.90&91.73&93.23\\ 
152&4.437075&0.000481&51.355038&0.005565&0.00023&0.00003&0.4883&0.1364&3.3&98.90&91.72&93.23\\ 
153&0.674955&0.000478&7.811985&0.005533&0.00023&0.00003&0.1507&0.1356&3.3&98.90&91.70&93.23\\ 
154&7.295163&0.000468&84.434760&0.005421&0.00024&0.00003&-0.1483&0.1329&3.5&98.91&91.69&93.24\\ 
155&8.289375&0.000476&95.941836&0.005509&0.00023&0.00003&-0.4974&0.1350&3.6&98.91&91.68&93.24\\ 
156&13.790030&0.000479&159.606828&0.005547&0.00023&0.00003&0.3534&0.1360&4.3&98.91&91.67&93.24\\ 
157&2.790366&0.000486&32.295906&0.005625&0.00023&0.00003&-0.3009&0.1379&3.3&98.92&91.66&93.24\\ 
158&6.330152&0.000466&73.265645&0.005390&0.00023&0.00003&0.0448&0.1321&3.3&98.92&91.65&93.24\\ 
159&11.546618&0.000482&133.641409&0.005580&0.00023&0.00003&0.1608&0.1368&3.9&98.92&91.64&93.24\\ 
160&10.219603&0.000484&118.282442&0.005602&0.00022&0.00003&-0.1732&0.1373&3.7&98.93&91.63&93.25\\ 
161&8.168456&0.000483&94.542314&0.005595&0.00022&0.00003&0.4513&0.1371&3.6&98.93&91.62&93.25\\ 
162&0.658442&0.000479&7.620860&0.005549&0.00023&0.00003&-0.2895&0.1360&3.2&98.93&91.61&93.25\\ 
163&16.161520&0.000492&187.054627&0.005696&0.00022&0.00003&0.4617&0.1396&5.3&98.94&91.60&93.25\\ 
164&7.442907&0.000462&86.144753&0.005352&0.00023&0.00003&-0.0362&0.1312&3.4&98.94&91.59&93.25\\ 
165&11.426610&0.000493&132.252429&0.005705&0.00022&0.00003&0.1803&0.1398&3.8&98.94&91.58&93.26\\ 
166&1.587122&0.000487&18.369468&0.005632&0.00022&0.00003&0.4358&0.1381&3.3&98.95&91.57&93.26\\ 
167&1.767490&0.000488&20.457055&0.005645&0.00022&0.00003&-0.2663&0.1384&3.3&98.95&91.56&93.26\\ 
168&1.666735&0.000489&19.290919&0.005656&0.00022&0.00003&-0.4967&0.1386&3.3&98.95&91.55&93.26\\ 
169&13.883301&0.000486&160.686356&0.005626&0.00022&0.00003&-0.0029&0.1379&4.2&98.96&91.54&93.26\\ 
170&10.390805&0.000456&120.263944&0.005273&0.00023&0.00003&0.3919&0.1292&3.7&98.96&91.53&93.27\\ 
171&12.741223&0.000485&147.467860&0.005615&0.00022&0.00003&-0.2775&0.1376&4.0&98.96&91.52&93.27\\ 
172&0.309166&0.000453&3.578306&0.005237&0.00023&0.00003&0.0626&0.1284&3.2&98.97&91.51&93.27\\ 
173&4.745479&0.000489&54.924531&0.005663&0.00022&0.00003&0.3323&0.1388&3.2&98.97&91.50&93.27\\ 
174&11.692495&0.000486&135.329801&0.005627&0.00022&0.00003&-0.0403&0.1379&3.8&98.97&91.49&93.27\\ 
175&1.359556&0.000488&15.735597&0.005650&0.00022&0.00003&0.4042&0.1385&3.3&98.97&91.48&93.28\\ 
176&0.975971&0.000488&11.295957&0.005645&0.00022&0.00003&-0.4811&0.1384&3.2&98.98&91.47&93.28\\ 
177&8.418358&0.000461&97.434701&0.005331&0.00023&0.00003&-0.2797&0.1307&3.5&98.98&91.46&93.28\\ 
178&9.786730&0.000488&113.272343&0.005650&0.00021&0.00003&-0.1297&0.1385&3.6&98.98&91.45&93.28\\ 
179&1.375485&0.000476&15.919966&0.005515&0.00022&0.00003&0.4005&0.1352&3.2&98.99&91.44&93.29\\ 
180&1.177101&0.000482&13.623856&0.005584&0.00021&0.00003&-0.2543&0.1369&3.2&98.99&91.43&93.29\\ 
181&5.267010&0.000497&60.960760&0.005752&0.00021&0.00003&-0.2691&0.1410&3.2&98.99&91.42&93.29\\ 
182&0.326968&0.000468&3.784349&0.005418&0.00022&0.00003&0.0205&0.1328&3.1&99.00&91.41&93.29\\ 
183&6.898981&0.000499&79.849312&0.005781&0.00021&0.00003&0.4291&0.1417&3.3&99.00&91.40&93.30\\ 
184&8.457694&0.000494&97.889973&0.005717&0.00021&0.00003&-0.3974&0.1401&3.4&99.00&91.39&93.30\\ 
185&10.319399&0.000499&119.437493&0.005773&0.00021&0.00003&0.3013&0.1415&3.5&99.00&91.38&93.30\\ 
186&9.755457&0.000500&112.910384&0.005792&0.00021&0.00003&-0.0400&0.1420&3.5&99.01&91.38&93.30\\ 
187&4.589204&0.000504&53.115787&0.005829&0.00021&0.00003&-0.3857&0.1429&3.1&99.01&91.37&93.31\\ 
188&6.132605&0.000488&70.979230&0.005648&0.00021&0.00003&-0.3332&0.1384&3.2&99.01&91.36&93.31\\ 
189&9.860127&0.000485&114.121844&0.005615&0.00021&0.00003&-0.0811&0.1376&3.5&99.02&91.35&93.31\\ 
190&9.924674&0.000499&114.868909&0.005773&0.00020&0.00003&0.4594&0.1415&3.5&99.02&91.34&93.32\\ 
191&1.029558&0.000447&11.916180&0.005176&0.00023&0.00003&-0.2873&0.1269&3.1&99.02&91.33&93.32\\ 
192&5.751525&0.000499&66.568575&0.005781&0.00020&0.00003&0.3417&0.1417&3.1&99.02&91.32&93.32\\ 
193&0.123769&0.000475&1.432511&0.005500&0.00022&0.00003&0.2788&0.1348&3.0&99.03&91.31&93.33\\ 
194&15.287241&0.000512&176.935655&0.005925&0.00020&0.00003&-0.1422&0.1452&4.5&99.03&91.31&93.33\\ 
195&7.389585&0.000473&85.527608&0.005471&0.00021&0.00003&0.2971&0.1341&3.2&99.03&91.30&93.34\\ 
196&11.619288&0.000500&134.482497&0.005789&0.00020&0.00003&0.4728&0.1419&3.6&99.03&91.29&93.34\\ 
197&6.229125&0.000508&72.096349&0.005883&0.00020&0.00003&0.0977&0.1442&3.1&99.04&91.28&93.34\\ 
198&7.351882&0.000499&85.091228&0.005776&0.00020&0.00003&0.2438&0.1416&3.2&99.04&91.27&93.35\\ 
199&7.285282&0.000476&84.320395&0.005512&0.00021&0.00003&-0.1116&0.1351&3.2&99.04&91.26&93.35\\ 
200&2.227274&0.000506&25.778639&0.005852&0.00020&0.00003&-0.2862&0.1434&3.1&99.05&91.25&93.35\\ 
\end{longtable}}

\newpage

\begin{table*}[htb]
\caption{\label{tbl:fixed_combos}Frequency table with fixed combinations between
parent and daughter modes. All AICs and BICs are calculated with respect to the
best model with only the first harmonic of the dominant mode.}
\begin{tabular}{rrrrrrrrrrrr}\hline\hline
$\Delta A/\bar{F}$&$f$ (d$^{-1}$)&$f$ ($\mu$Hz)&$\phi$&$\log_{10}(A_r)$&$\epsilon_{\log_{ 10}(A_r)}$&$\phi_r$&$e_{\phi_r}$&SNR&VR&AIC&BIC\\ \hline
0.036966&5.486888&63.50565&-0.0355   &-       &-      &-    &   -&188.7&91.86&100.00&100.01\\ 
0.006313&10.973776&127.01130&-0.4967 &-       &-      &-    &   -&63.0&94.54&98.25&98.20\\ 
0.004049&16.460665&190.51695&0.3490  &-       &-      &-    &-   &70.4&95.64&97.26&97.18\\ 
0.003125&0.299178&3.46271&-0.3753    &-       &-      &-    &-   &27.8&96.31&96.54&96.45\\ 
0.002220&6.324787&73.20355&-0.2322   &-       &-      &-    &-   &21.8&96.64&96.12&96.04\\ 
0.001568&8.408990&97.32628&0.4746    &-       &-      &-    &-   &17.9&96.81&95.91&95.83\\ 
0.001441&7.255010&83.97002&-0.0764   &-       &-      &-    &-   &16.5&96.95&95.72&95.65\\ 
0.001269&21.947553&254.02260&0.1169  &-       &-      &-    &-   &36.6&97.06&95.56&95.51\\ 
0.001181&11.811675&136.70920&-0.4118 &-4.82   &0.01   &-0.14&0.01&15.4&97.15&95.42&95.36\\ 
0.001003&6.143123&71.10096&0.0079    &-       &-      &-    &-   &10.1&97.22&95.32&95.28\\ 
0.000999&7.103595&82.21753&0.3435    &-       &-      &-    &-   &11.5&97.29&95.21&95.19\\ 
0.000924&1.111887&12.86906&-0.0844   &-3.17   &0.02   &-0.06&0.02&9.4&97.35&95.11&95.09\\ 
0.000936&13.895879&160.83193&0.2930  &-4.76   &0.02   &-0.12&0.01&14.9&97.41&95.01&94.98\\ 
0.000852&7.358900&85.17245&0.3698    &-       &-      &-    &-   &10.2&97.46&94.94&94.92\\ 
0.000848&1.050091&12.15383&0.2169    &-3.17   &0.02   & 0.17&0.02&8.6&97.50&94.85&94.84\\ 
0.000826&0.898758&10.40229&0.1134    &-       &-      &-    &-   &8.5&97.55&94.78&94.77\\ 
0.000644&11.630011&134.60661&-0.1515 &-4.74   &0.02   &-0.14&0.02&8.9&97.58&94.72&94.72\\ 
0.000653&0.837898&9.69790&-0.4485    &-5.07   &0.02   &-0.25&0.02&6.7&97.61&94.67&94.67\\ 
0.000579&2.922102&33.82063&-0.4912   &-4.98   &0.02   &0.06 &0.02&6.0&97.63&94.63&94.62\\ 
0.000529&8.770800&101.51388&-0.0181  &-       &-      &-    &-   &6.9&97.65&94.60&94.61\\ 
0.000448&5.187710&60.04294&0.4809    &-5.39   &0.02   &0.13 &0.02&4.9&97.66&94.58&94.59\\ 
0.000426&10.674598&123.54859&-0.2978 &-5.41   &0.03   &0.40 &0.02&5.9&97.68&94.55&94.56\\ 
0.000424&10.375420&120.08588&-0.2473 &-5.41   &0.03   &0.12 &0.02&5.7&97.69&94.53&94.54\\ 
0.000345&8.366896&96.83908&0.1436    &-3.60   &0.04   &0.10 &0.03&5.5&97.70&94.51&94.52\\ 
0.000390&6.265160&72.51343&0.0363    &-       &-      &-    &-   &4.7&97.71&94.50&94.53\\ 
0.000393&17.298563&200.21485&0.3194  &-5.30   &0.03   &-0.36&0.02&9.1&97.72&94.48&94.51\\ 
0.000363&4.648990&53.80775&0.1698    &-5.33   &0.03   &0.05 &0.02&4.2&97.73&94.46&94.49\\ 
0.000339&9.799869&113.42441&0.1004   &-3.36   &0.04   &0.38 &0.04&4.9&97.73&94.45&94.47\\ 
0.000289&5.786066&66.96836&0.2604    &-5.58   &0.03   &-0.37&0.02&3.9&97.74&94.44&94.46\\ 
0.000226&14.257688&165.01953&-0.1453 &-4.91   &0.04   &-0.06&0.03&4.6&97.74&94.43&94.45\\ 
0.000234&11.752049&136.01908&-0.1853 &-4.77   &0.05   &-0.16&0.03&4.0&97.75&94.42&94.45\\ 
0.000205&12.845788&148.67810&0.0760  &-5.16   &0.04   &-0.17&0.03&4.2&97.75&94.42&94.44\\ 
0.000206&4.589353&53.11751&-0.3901   &-5.72   &0.04   &-0.47&0.03&3.1&97.75&94.41&94.44\\ 
\hline\end{tabular}
\end{table*}

\newpage

\begin{table}[htb]
\caption{\label{tbl:var_3}Variable amplitude/phase model: parameters for $a_1$. $C=0.03705$}
\begin{tabular}{rrrrrrr}\hline\hline
$\Delta A/\bar{F}$&$\epsilon_A$&$f$ (d$^{-1}$)&$\epsilon_f$&$f$ ($\mu$Hz)&$\phi$&$\epsilon_phi$\\ \hline
0.00244&0.00011&0.8379&0.0002&9.6985&0.13&0.05\\ 
0.00117&0.00011&1.7679&0.0003&20.4613&0.41&0.10\\ 
0.00087&0.00011&0.6564&0.0004&7.5976&0.34&0.10\\ 
0.00075&0.00011&1.7588&0.0005&20.3564&-0.39&0.10\\ 
0.00066&0.00011&1.6160&0.0006&18.7031&-0.18&0.20\\ 
0.00064&0.00011&0.2999&0.0006&3.4708&-0.45&0.20\\ 
0.00062&0.00010&1.8717&0.0006&21.6636&-0.17&0.20\\ 
0.00051&0.00010&1.8870&0.0007&21.8400&0.23&0.20\\ 
0.00046&0.00010&0.1976&0.0007&2.2865&0.21&0.20\\ 
0.00045&0.00010&1.9380&0.0007&22.4302&-0.41&0.20\\ 

\hline\end{tabular}
\end{table}

\begin{table}[htb]
\caption{\label{tbl:var_4}Variable amplitude/phase model: parameters for $a_2$. $C=0.00675$}
\begin{tabular}{rrrrrrr}\hline\hline
$\Delta A/\bar{F}$&$\epsilon_A$&$f$ (d$^{-1}$)&$\epsilon_f$&$f$ ($\mu$Hz)&$\phi$&$\epsilon_phi$\\ \hline
0.00131&0.00011&2.5651&0.0003&29.6883&0.42&0.08\\ 
0.00085&0.00010&0.8377&0.0004&9.6958&0.34&0.10\\ 
0.00075&0.00010&1.7678&0.0005&20.4611&0.37&0.10\\ 
0.00061&0.00010&1.6166&0.0005&18.7106&-0.19&0.20\\ 
0.00055&0.00010&0.5992&0.0006&6.9349&0.04&0.20\\ 
0.00047&0.00010&1.8708&0.0007&21.6530&-0.06&0.20\\ 
0.00046&0.00008&0.6130&0.0007&7.0951&0.14&0.20\\ 
0.00045&0.00008&2.6428&0.0007&30.5880&0.34&0.20\\ 
0.00045&0.00008&0.2991&0.0007&3.4612&0.03&0.20\\ 
0.00043&0.00008&0.6561&0.0007&7.5937&-0.44&0.20\\ 

\hline\end{tabular}
\end{table}

\begin{table}[htb]
\caption{\label{tbl:var_5}Variable amplitude/phase model: parameters for $a_3$. $C=0.00442$}
\begin{tabular}{rrrrrrr}\hline\hline
$\Delta A/\bar{F}$&$\epsilon_A$&$f$ (d$^{-1}$)&$\epsilon_f$&$f$ ($\mu$Hz)&$\phi$&$\epsilon_phi$\\ \hline
0.00087&0.00006&2.5650&0.0002&29.6870&0.46&0.07\\ 
0.00049&0.00005&0.8376&0.0004&9.6942&0.22&0.10\\ 
0.00027&0.00005&0.5993&0.0007&6.9367&-0.02&0.20\\ 
0.00026&0.00005&2.6425&0.0007&30.5840&0.42&0.20\\ 
0.00026&0.00005&0.6569&0.0008&7.6031&0.34&0.20\\ 
0.00026&0.00005&1.7681&0.0007&20.4638&0.37&0.20\\ 
0.00024&0.00005&0.6126&0.0008&7.0899&0.16&0.20\\ 
0.00024&0.00005&2.2028&0.0008&25.4959&-0.12&0.20\\ 
0.00023&0.00005&2.3188&0.0008&26.8375&-0.07&0.20\\ 
0.00023&0.00005&0.2985&0.0008&3.4552&0.14&0.20\\ 

\hline\end{tabular}
\end{table}

\begin{table}[htb]
\caption{\label{tbl:var_6}Variable amplitude/phase model: parameters for $a_4$. $C=0.00203$}
\begin{tabular}{rrrrrrr}\hline\hline
$\Delta A/\bar{F}$&$\epsilon_A$&$f$ (d$^{-1}$)&$\epsilon_f$&$f$ ($\mu$Hz)&$\phi$&$\epsilon_phi$\\ \hline
0.00016&0.00004&0.8375&0.0008&9.6938&-0.45&0.20\\ 

\hline\end{tabular}
\end{table}

\begin{table}[htb]
\caption{\label{tbl:var_7}Variable amplitude/phase model: parameters for $\phi_1$. $C=0.00203$}
\begin{tabular}{rrrrrrr}\hline\hline
$\Delta A/\bar{F}$&$\epsilon_A$&$f$ (d$^{-1}$)&$\epsilon_f$&$f$ ($\mu$Hz)&$\phi$&$\epsilon_phi$\\ \hline
0.00835&0.00050&0.8378&0.0002&9.6968&-0.13&0.07\\ 
0.00539&0.00050&1.7679&0.0003&20.4616&0.12&0.10\\ 
0.00482&0.00050&0.6564&0.0004&7.5976&0.07&0.10\\ 
0.00436&0.00050&2.5644&0.0004&29.6802&-0.15&0.10\\ 
0.00374&0.00040&1.6165&0.0004&18.7092&-0.48&0.10\\ 
0.00344&0.00040&1.8718&0.0004&21.6646&-0.44&0.10\\ 
0.00306&0.00040&1.7588&0.0005&20.3563&0.30&0.10\\ 
0.00234&0.00040&1.8876&0.0006&21.8470&-0.11&0.20\\ 
0.00200&0.00040&0.7411&0.0007&8.5776&0.24&0.20\\ 
0.00197&0.00040&0.1982&0.0007&2.2940&-0.08&0.20\\ 

\hline\end{tabular}
\end{table}

\begin{table}[htb]
\caption{\label{tbl:var_8}Variable amplitude/phase model: parameters for $\phi_2$. $C=0.00203$}
\begin{tabular}{rrrrrrr}\hline\hline
$\Delta A/\bar{F}$&$\epsilon_A$&$f$ (d$^{-1}$)&$\epsilon_f$&$f$ ($\mu$Hz)&$\phi$&$\epsilon_phi$\\ \hline
0.02791&0.00300&0.8380&0.0004&9.6994&0.11&0.10\\ 
0.02607&0.00300&2.5646&0.0004&29.6829&-0.34&0.10\\ 
0.02034&0.00300&0.6572&0.0005&7.6063&0.33&0.10\\ 
0.01615&0.00300&0.6111&0.0006&7.0729&0.43&0.20\\ 
0.01317&0.00300&1.7676&0.0007&20.4588&0.20&0.20\\ 
0.01133&0.00300&0.2992&0.0008&3.4624&0.29&0.20\\ 
0.01123&0.00200&1.8724&0.0008&21.6711&-0.40&0.20\\ 
0.01044&0.00200&1.1135&0.0008&12.8875&0.18&0.20\\ 
0.01073&0.00200&1.7594&0.0008&20.3632&0.31&0.20\\ 
0.01022&0.00200&0.5977&0.0008&6.9179&0.19&0.20\\ 

\hline\end{tabular}
\end{table}

\begin{table}[htb]
\caption{\label{tbl:var_9}Variable amplitude/phase model: parameters for $\phi_3$. $C=0.00203$}
\begin{tabular}{rrrrrrr}\hline\hline
$\Delta A/\bar{F}$&$\epsilon_A$&$f$ (d$^{-1}$)&$\epsilon_f$&$f$ ($\mu$Hz)&$\phi$&$\epsilon_phi$\\ \hline
0.03866&0.00300&2.5646&0.0003&29.6825&-0.33&0.08\\ 
0.02706&0.00300&0.8379&0.0004&9.6982&-0.01&0.10\\ 
0.01324&0.00300&0.6571&0.0008&7.6055&0.25&0.20\\ 
0.01300&0.00300&1.7679&0.0008&20.4618&0.28&0.20\\ 
0.01270&0.00300&1.9803&0.0008&22.9200&0.17&0.20\\ 
0.01199&0.00300&0.6095&0.0008&7.0547&0.50&0.20\\ 
0.01183&0.00300&2.0427&0.0008&23.6424&-0.41&0.20\\ 
0.01106&0.00300&2.5777&0.0009&29.8348&-0.05&0.30\\ 
0.01101&0.00300&1.1718&0.0009&13.5625&0.16&0.30\\ 
0.01049&0.00300&2.1939&0.0009&25.3920&-0.49&0.30\\ 

\hline\end{tabular}
\end{table}

\begin{table}[htb]
\caption{\label{tbl:var_10}Variable amplitude/phase model: parameters for $\phi_4$. $C=0.00203$}
\begin{tabular}{rrrrrrr}\hline\hline
$\Delta A/\bar{F}$&$\epsilon_A$&$f$ (d$^{-1}$)&$\epsilon_f$&$f$ ($\mu$Hz)&$\phi$&$\epsilon_phi$\\ \hline
0.05623&0.00800&0.8378&0.0005&9.6970&0.16&0.10\\

\hline\end{tabular}
\end{table}
\end{appendix}
\end{document}